\documentclass{article}

\usepackage[dblblindworkshop, final]{neurips_2026}
\workshoptitle{Sim2Science: ML with Imperfect Scientific Models}

\usepackage[utf8]{inputenc}
\usepackage[T1]{fontenc}
\usepackage{float} %
\usepackage{hyperref}
\hypersetup{colorlinks=true, linkcolor=black,
  citecolor=black, urlcolor=black}
\usepackage{url}
\usepackage{booktabs}
\usepackage{amsfonts,amsmath,amssymb}
\usepackage{nicefrac}
\usepackage{microtype}
\usepackage[dvipsnames]{xcolor}
\usepackage{graphicx}
\usepackage[export]{adjustbox}
\usepackage{comment}
\newcommand{\scanpanel}{0.48\linewidth} %
\newcommand{\scangap}{\hfill} %
\usepackage{placeins} %

\graphicspath{{fig/}}

\newcommand{\cmark}{\checkmark}
\newcommand{\xmark}{$\times$}
\newcommand{\modelname}{ModeCast}

\makeatletter
\renewcommand{\paragraph}{\@startsection{paragraph}{4}{\z@}%
  {0pt}{-1em}{\normalsize\bfseries}}
\makeatother

\title{\modelname: One-Shot Prediction of Linear MHD Eigenmodes and Growth Rates in Tokamaks}

\author{%
  Alex Higginbottom \quad Zara Z.~Ercan \quad Abetharan Antony \quad Guzman Sanchez Gonzalez \\
  \textbf{Rafael Grazzini Placucci \quad Armand Kassa\"{\i} Koupa\"{\i} \quad
  Artur P.~Toshev} \\[0.4em]
  Zenithon \\
}

\begin{document}

\maketitle

\begin{abstract}
Large-scale magnetohydrodynamic instabilities break plasma confinement, can trigger disruptions, and constrain fusion reactor design; modeling a single configuration against them takes hours with a reference solver such as JOREK, and a design study requires thousands of hours. \modelname\ is the first learned surrogate to return the complete linear stability assessment for MHD modes. Trained on a comprehensive dataset of over 45,000 JOREK simulations across thousands of equilibria, it screens stability across toroidal mode numbers $n{=}1$--$4$ and predicts the full spatial structure and growth rate of unstable modes. A second network of the same architecture can replace the equilibrium solve too, making the full pipeline up to $10^6\times$ faster than JOREK on one GPU. The gain comes from building the linear-phase physics into the formulation, objective, and inputs, yielding up to six times more training data. \modelname\ misses none of the unstable held-out cases, and its predicted fields reproduce the simulation's measured physics, from magnetic island topology to resistivity trends.{} On a fixed-geometry resistivity scan outside its training distribution, refitting only the 22k-parameter growth rate head of the 145M-parameter network on four simulation halves the held-out internal kink growth rate error, and fine-tuning the whole network adds little.{} Because it is differentiable and returns full fields, \modelname\ could in the future assimilate experimental measurements, providing a bridge to real devices and accelerating progress toward commercial fusion. Project page: \href{https://zenithon.ai/research/modecast}{zenithon.ai/research/modecast}.
\end{abstract}

\section{Introduction}
\label{sec:intro}
Tokamaks are the leading confinement concept for fusion power, and large-scale magnetohydrodynamic (MHD) instabilities, growing from small perturbations of the plasma equilibrium, degrade confinement and set hard operational limits on any design~\citep{zhu2023fishbone}. Assessing a configuration's stability is a central, recurring task, and the reference method is expensive: initialize a nonlinear MHD code such as JOREK~\citep{hoelzl2021jorek, pamela2020fullmhd, artola2021vde}, perturb, and evolve for hours on tens of cores per linear run, for thousands of configurations and modes.

\begin{figure}[t]
  \centering
  \includegraphics[width=0.76\linewidth]{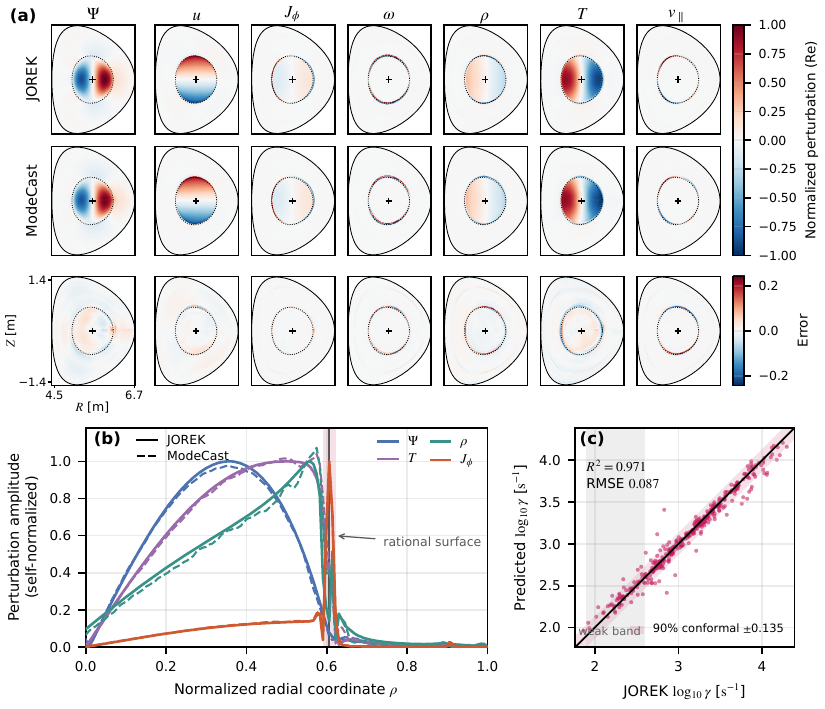}
  \caption{One forward pass predicts the seven-field complex eigenmode and
  growth rate; the full time-dependent perturbation reconstructs the full state. (a) Truth, prediction, and error for an unseen configuration, shared per-channel normalization. (b) Radial amplitude profiles, resonant $q{=}1$ surface, and measured layer band. (c) Growth rates with 90\% confidence intervals.}
  \label{fig:showcase}
\end{figure}

Existing methods trade speed against completeness. Reduced models and analytic criteria are fast but partial~\citep{haverkort2016fmhd}. Learned scalar surrogates, trained one instability family at a time, return only a growth rate or a stability label, leaving the identity and location of the mode unknown~\citep{piccione2020pb, bruncrona2025pb, xu2024dtm, liu2024fishbone, kumar2026vertical}. Field-level surrogates resolve the fields, but target neighboring regimes and forecast evolution in time rather than one-shot linear stability~\citep{carey2025fno, paischer2025gyroswin, gopakumar2023fno}. The closest prior work~\citep{chang2025dltm} covers a single-mode family in cylindrical geometry and reports accuracy only on scalar targets. No existing surrogate answers the question a designer asks of a configuration: which mode is unstable, where it sits, and how fast it grows.

\modelname\ is, to our knowledge, the first learned system to return that assessment, covering toroidal mode numbers $n{=}1$--$4$ across 12{,}990 shaped equilibria, with seven spatially resolved fields per mode (Figure~\ref{fig:showcase}) and, through the factorization, the full time evolution of the linear phase. \modelname\ is not restricted to one instability family by construction: each run fixes $n$ and the equilibrium, and the fastest-growing mode emerges from the resistive MHD solution whatever its type, so we characterize each mode by its $n$, resonant surface $q{=}m/n$, and parity (Appendix~\ref{app:physics}; Table~\ref{tab:composition}).

Our contributions are threefold. (i) A completely learned stability assessment: we formulate linear stability analysis as one-shot regression from the equilibrium to the mode shape and growth rate, and build the first system that delivers it at scale, with a calibrated classifier that abstains for outcome reliability, a forecaster for the full seven-field eigenmode, and, through the factorization, the exact linear time evolution. (ii) A physics-conditioned training recipe: the factorization fixes the eigenmode's gauge, a single global phase whose repair cuts physical-gauge error by 67\% (Section~\ref{sec:method}), and safety-factor features the equilibrium solver already computes raise accuracy and data efficiency by amounts we measure in equivalent training data (Section~\ref{sec:results}). (iii) An evaluation protocol for near-singular eigenmodes, norms in two gauges, structural similarity, and recovered physics properties, under which a surrogate's predictions are verified as physics rather than pixels (Section~\ref{sec:results}).
{} Finally, the trained representation transfers: on a fixed-geometry resistivity scan outside the training distribution, refitting the 22k-parameter growth rate head on four runs halves the held-out internal kink growth rate error (Section~\ref{sec:transfer}).{}

\section{Task and data}
\label{sec:task}
\paragraph{Task.}
A linear JOREK run perturbs a shaped tokamak equilibrium (a stationary MHD solution) with a single toroidal harmonic $n$ and evolves seven coupled perturbation fields on the poloidal plane, the vertical cross-section of the torus (Appendix~\ref{app:geometry&symbols}): magnetic flux $\Psi$, velocity stream function $u$, toroidal current density $J_\phi$, vorticity $\omega$, density $\rho$, temperature $T$, and parallel velocity $v_\parallel$. Each case needs an equilibrium solve (about 2~min on 32 cores) and a linear run (1.3--4.3~h); \modelname\ replaces the linear run, and a second network of the same architecture can replace the solve too (Appendix~\ref{app:e2e}). JOREK discretizes the poloidal plane in $ G^1$-continuous B\'ezier finite elements and the toroidal direction via a Fourier series \citep{hoelzl2021jorek}. Within a quadrangular element $e$, and omitting the toroidal Fourier expansion for clarity, any physical variable $f$ is represented as
\begin{equation}
   f(s_1,s_2)|_e = \sum_{k=1}^{n_{\text{vert}}} \sum_{l=1}^{n_{\text{dofs}}}
   f_{k,l}\, H_{k,l}(s_1,s_2)\, d^{e}_{k,l},
   \label{eq:jorek_expansion}
\end{equation}
where $H_{k,l}$ are 2D poloidal basis functions in element-local coordinates $(s_1,s_2)$, $d^{e}_{k,l}$ are element-local scale factors, and the complex coefficients $f_{k,l}$ are stored as $\cos(n\varphi)$ and $\sin(n\varphi)$ pairs. A linear run holds the equilibrium fixed and evolves a perturbation $\delta f_n$ in a single $n$; once the initial transient decays, the dominant eigenmode sets the trajectory and all time dependence sits in the coefficients, which grow exponentially
  \begin{equation}
    \delta f_{k,l}(t) = \hat f_{k,l}\, e^{\gamma t}
    \quad\Longrightarrow\quad
    \delta f_n(R,Z,\varphi,t)
      = \mathrm{Re}\!\left[\hat f_n(R,Z)\, e^{i n\varphi}\, e^{\gamma t}\right],
    \label{eq:ansatz}
  \end{equation}
with $\hat{f}_n|_e = \sum_{k,l} \hat{f}_{k,l} H_{k,l}\, d^{e}_{k,l}$ the complex mode shape assembled from the same basis as Equation~\eqref{eq:jorek_expansion}, and $\gamma$ the growth rate \citep{haverkort2016fmhd}; if $\gamma \le 0$ the plasma is linearly stable against the perturbation. Predicting the trajectory is then equivalent to predicting $(\hat f_n, \gamma)$: toroidal harmonics decouple about an axisymmetric equilibrium, so one complex 2D field per variable plus one scalar reconstructs the full 3D, time-dependent perturbation over the linear phase, with no rollout to compound error (Table~\ref{tab:capabilities}, Appendix~\ref{app:systems}). The factorization also fixes the gauge freedom that Section~\ref{sec:results} quantifies: $\hat f_n$ is defined up to one global phase, while the relative phases between fields are physical.

\paragraph{Resistive scaling of the growth rate.}
  Tearing modes and the resistive internal kink grow by reconnecting the magnetic field in a thin layer around the rational surface $q = m/n$, so their growth rates depend on the resistivity $\eta$. For large Lundquist number and negligible viscosity, theory gives $\gamma \propto \eta^{3/5}$ for the $m{=}2$, $n{=}1$ tearing mode \citep{furth1963tearing} and $\gamma \propto \eta^{1/3}$ for the $m{=}n{=}1$ resistive internal kink \citep{coppi1976kink}.

\paragraph{Data.}
Six campaigns of linear JOREK simulations yield 46{,}034 (equilibrium, $n$) pairs over 12{,}990 equilibria with $n = 1$--$4$, split so that no geometry appears in more than one of train, validation, and test. Two training sets are drawn from this pool: a curated \emph{small dataset} of 2{,}718 runs and a \emph{large dataset} of 9{,}968 runs admitted under a relaxed fit-quality floor (Appendix~\ref{app:dataset}).

\section{Method}
\label{sec:method}

\modelname\ has two learned components: a stability classifier that screens inputs, calibrated so that borderline cases abstain rather than pass, and a forecaster network that predicts the eigenmode and growth rate for the remainder. The system is agnostic to its backbone network; the instantiation throughout is a 144.9M-parameter convolutional U-Net~\citep{ronneberger2015unet} (backbone comparison in Section~\ref{sec:results}).

\paragraph{Eigenmode and growth rate prediction.}
For inputs identified unstable, the network maps the $t{=}0$ equilibrium to $(\hat f_n, \log_{10}\gamma)$. The baseline inputs are the four equilibrium field channels ($\Psi, J_\phi, \rho, T$), two coordinate maps (R, Z), the design parameters, and the mode number. On top of these sit the physics features this paper adds: two feature maps derived from the equilibrium's safety-factor profile (the $q$-profile map and the signed proximity to its rational surfaces, where resistive modes localize) and conditioning scalars comprising $q$-profile summaries, three dimensionless similarity groups (Lundquist number, magnetic Prandtl number, normalized pressure), and the Alfv\'en time: 8 input field channels and 39 scalars in all.
Every input is computed during the equilibrium solve that precedes each stability run, and each set is priced in Section~\ref{sec:results}.

\paragraph{Training objective and gauges.}
The loss for complex mode shapes must be phase-invariant, and per Equation~\eqref{eq:ansatz} the only admissible invariance is one phase shared by all channels. Our objective normalizes per channel and aligns the single shared phase before penalizing errors, so that the $C{-}1$ physical relative phases between fields are trained; the standard per-channel phase-invariant loss, which leaves them free, serves as the phase-blind control in Section~\ref{sec:results}.
Evaluation mirrors the two choices: the \emph{physical gauge} (one shared phase, joint relative $L_2$) scores the relative phases as errors and decides all model selection; the conventional \emph{per-channel gauge} is reported alongside.

\paragraph{Controlled resistivity scan.}
\label{sec:scan}
Our training design samples resistivity using a Latin hypercube alongside 20 other parameters, drawing one value per equilibrium. Geometry is shared across the modes run on a given equilibrium, but no two runs share a geometry at different $\eta$, so the dataset supports marginal trends in $\eta$ and not exponents (Appendix~\ref{app:physics}). To test the resistivity response at fixed geometry, we build an external scan following the reduced-MHD resistivity benchmark of \citet{pamela2020fullmhd}: a large-aspect-ratio, circular, low-$\beta$ tokamak perturbed at $n{=}1$, run as two branches, a $1/1$ internal kink and a $2/1$ tearing mode, each swept in $\eta$ at the fixed geometry setting of these scalings. The published inputs were not available, so we reconstructed the equilibrium from the published description and generated the reference growth rates with our own JOREK runs using model 600, the reduced-MHD model that produced our training data, at the benchmark's diffusive settings, $\mu = 0$ for the kink and $\mu = 10^{-8}$ for the tearing case. The surrogate is scored against these reference runs throughout; the published points, computed with JOREK model 303, appear in the figures for reference only. Two features place the case outside our training distribution: the cross-section is circular, whereas our equilibria are D-shaped, and the kink case carries no viscosity. All values of $\eta$ and $\mu$ are in JOREK-normalized units.

\paragraph{Training window.}
The kink scan covers $\eta = 10^{-10}$ to $10^{-2}$ and the tearing scan $10^{-9}$ to $2\times10^{-5}$, both wider than the range our model was trained on. Training samples the Lundquist number log-uniformly over $S = 10^{6}$ to $10^{9}$, which across our equilibria corresponds to $\eta = 1.293\times10^{-9}$ to $2.394\times10^{-5}$. Runs inside this window are called native, and runs outside it are out-of-distribution. Every exponent fit uses the native runs only, five per branch; the few-shot refits below draw their training runs from the whole scan and report native and out-of-distribution error separately. We quote all windows in $\eta$ rather than $S$, because $S$ at fixed $\eta$ depends on geometry and density, and the benchmark's circular equilibrium does not share ours.

\paragraph{Few-shot adaptation.}
Each branch of the scan is its own few-shot problem, with 11 internal kink runs and 7 tearing runs. We draw 200 random permutations of each branch's $N$ runs; shot $k$ trains on the first $k$ runs of a permutation and is scored on the remaining $N-k$. We freeze the 145M-parameter trunk and refit only the 22k-parameter growth rate head from its pretrained weights, minimizing mean squared error on $\log_{10}\gamma$ with Adam (learning rate $10^{-3}$, 300 steps, fixed in advance). Each of the four ensemble members is refitted separately, and their predicted $\log_{10}\gamma$ values are averaged. We report the median over permutations of the mean absolute error (in dex) across the held-out runs. As references, the pretrained model without refitting and a two-parameter straight line in $\log_{10}\eta$ fitted to the same $k$ runs are scored on the same held-out sets. At $\mu = 0$, the kink branch's magnetic Prandtl number lies outside the training range, so we score two input sets: \emph{raw}, with the benchmark viscosity, and \emph{visco-fix}, with $\mu$ replaced by the training median $3.25\times10^{-6}$ in the network's inputs only. Visco-fix, therefore, describes a viscous layer that the reference simulation lacks; it measures how much of the out-of-distribution error a given input accounts for, not agreement at a matched viscosity.

\section{Results}
\label{sec:results}

All comparisons are paired on shared cases, with 95\% bootstrap confidence intervals (CIs), and model selection uses the validation split alone (details in Appendix~\ref{app:results}). Any single metric fails to characterize a near-singular eigenmode, so we evaluate at three levels: energy-weighted norms in two gauges, structural similarity (SSIM), and recovered physics properties.
\paragraph{Stability screening.}
The classifier runs first, resolving the stable majority of inputs and passing the remainder to the forecaster. At its conservative operating point, it misses none of the 1{,}258 unstable test cases (one-sided 95\% upper bound on the miss rate: 0.24\%; one miss in 1{,}344 on validation), at a false-alarm rate of 19.1\% per configuration (9.45\% per input; policy and example in Appendix~\ref{app:systems}).

\paragraph{Comparison with conventional training.}
Table~\ref{tab:main} compares five architectures under two training regimes: conventional training on bare fields and design scalars with a per-channel loss, and training within \modelname. Conventionally trained, every backbone remains close to the no-information regime set by the nearest-neighbor lookup; within the system, the same backbones reduce their error by more than half, and the difference between training regimes exceeds the difference between architectures several times over (Table~\ref{tab:ablation}). The training recipe, not the choice of backbone, determines performance.

\begin{table}[h]
  \caption{System-level comparison (validation, 358 cases; relative $L_2$
  medians; $\gamma$ RMSE in $\log_{10}$; test 355 cases, run once, --- where a row postdates it).
  Best bold, second underlined; 95\% CIs in Appendix~\ref{app:results}, every \modelname\ margin over the twins CI-separated.}
  \label{tab:main}
  \centering
  \resizebox{\linewidth}{!}{%
  \begin{tabular}{lcccccc}
    \toprule
    Method & Params & Rel.\ $L_2$ (physical) & (per-channel) & $\gamma$ $R^2$ & $\gamma$ RMSE & Test\\
    \midrule
    \multicolumn{7}{l}{\emph{Classical baselines}}\\
    $k$-NN lookup ($k{=}5$) & --- & 0.892 & 0.825 & 0.532 & 0.349 & 0.899\\
    \midrule
    \multicolumn{7}{l}{\emph{Conventional training (same parameter-matched backbones, bare inputs, per-channel loss)}}\\
    U-Net~\citep{ronneberger2015unet} & 145M & 0.761 & 0.682 & 0.823 & 0.215 & ---\\ %
    FNO~\citep{li2021fno} & 143M & 0.789 & 0.718 & 0.812 & 0.222 & ---\\ %
    ViT~\citep{dosovitskiy2021vit} & 156M & 0.865 & 0.761 & 0.744 & 0.259 & ---\\ %
    UPT~\citep{alkin2024upt} & 190M & 0.747 & 0.660 & 0.876 & 0.180 & ---\\ %
    CALM-PDE~\citep{hagnberger2025calmpde} & 177M & 0.832 & 0.771 & 0.791 & 0.234 & ---\\ %
    \midrule
    \multicolumn{7}{l}{\emph{Ours (evaluated on the curated benchmark)}}\\
    \modelname\ (single, small dataset) & 145M & \underline{0.304} & \underline{0.252} & 0.968 & 0.091 & 0.322\\
    \textbf{\modelname\ (ensemble of 4, small dataset)} & 4$\times$145M & \textbf{0.285} {\scriptsize(0.261--0.304)} & \textbf{0.232} & \underline{0.971} & \underline{0.087} & \textbf{0.298}\\
    \modelname\ (single, large dataset) & 145M & 0.309 & 0.261 & \textbf{0.976} & \textbf{0.079} & \underline{0.318}\\ %
    \bottomrule
  \end{tabular}}
\end{table}

\paragraph{Physical versus per-channel gauge.}
The per-channel metric aligns each field's phase separately and so ignores the relative phases between fields, which are physical: the eigenmode carries only one global phase. The physical gauge counts them. The phase-blind control shows the gap: it matches the single \modelname\ model per channel yet scores nearly three times worse in the physical gauge, and it is the model whose field lines misplace the island chain (Figure~\ref{fig:poincare}).

\paragraph{Mode structure.}
Beyond the aggregate score, the ensemble (four members~\citep{lakshminarayanan2017ensembles}, complex-averaged after phase alignment; Appendix~\ref{app:results}) replicates on the once-run test split within its validation interval, and its error divides along physical lines (Table~\ref{tab:channels}; exemplars in Appendix~\ref{app:gallery}). The three smooth channels are recovered to within 5--14\%; the four that carry the resistive layer sit at 18--38\%, with a heavy tail concentrated within a few layer widths of the rational surface. The best-recovered channel, $\Psi$, is also the most consequential, since the field topology derives from it alone. The layer channels' residual is dominated by sub-cell placement of a near-discontinuous sheet: SSIM exceeds 0.9 on every channel, while displacing the \emph{true} $\omega$ field by a single grid cell, scored against itself, already reads 0.99 in relative $L_2$ (Figure~\ref{fig:shiftscan}). A best-case realignment search bounds the model's misregistration to 0.1--0.3 cells, about a fifth of the layer-channel error (Appendix~\ref{app:results}).

\begin{figure}[t]
  \centering
  \includegraphics[width=0.85\linewidth]{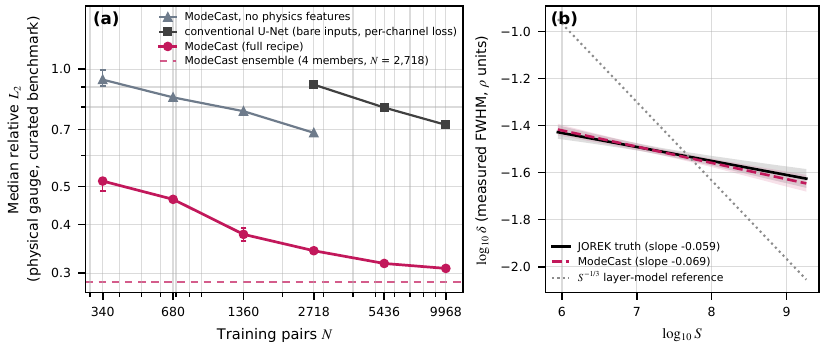}
  \caption{(a) Error--data scaling, same 145M U-Net backbone throughout: the full recipe runs 2--3$\times$ below the conventionally trained U-Net at every size; dashed line, four-member ensemble. (b) Measured resistive-layer width against Lundquist number, truth and prediction (95\% bands, 358 cases): the prediction carries the truth's trend; both floor near the grid cell (Appendix~\ref{app:results}).}
  \label{fig:scaling}
\end{figure}

\paragraph{Growth rates.}
The same picture holds for $\gamma$, reported as $R^2$ and RMSE in $\log_{10}\gamma$. The ensemble reaches $R^2 = 0.971$ on validation and 0.966 on the once-run test split, and its split-conformal intervals, the 90\% bands of Figure~\ref{fig:showcase}c, cover 91.0\% of test cases. Boosted trees on the 35 scalar features alone reach $R^2 = 0.85$, so the field surrogate halves the log-space error. Unlike the mode structure, $\gamma$ is limited by data rather than by representation: trained on the large dataset and scored on the same curated benchmark, the single model improves from RMSE 0.091 to 0.079 and posts the best once-run test $\gamma$ of any row in Table~\ref{tab:main} ($R^2$ 0.981, RMSE 0.071; counterfactual in Table~\ref{tab:counterfactual}).
\paragraph{Data efficiency and scaling.}
Physics conditioning also serves as a substitute for training data. Across a grid of four nested feature sets and five dataset sizes, error falls monotonically, and the sets stay ordered, so each feature increment can be priced as the extra data the poorer set needs to match it (Table~\ref{tab:prices}); in simulation cost, matching the conditioned model's growth rate accuracy with bare inputs would take roughly three times the small training set's compute again. The picture holds at scale: extended to 9{,}968 pairs, the conventionally trained twin improves faster in relative terms but from more than twice the error, and remains well above \modelname\ at every size measured (Figure~\ref{fig:scaling}a).

\paragraph{Reproducing key physics.}
Because the surrogate returns the full eigenmode, its physics can be read directly off the predictions and checked; verification that no scalar surrogate permits (protocols and CIs in Appendix~\ref{app:physics}). \emph{Operator structure:} predicted vorticity matches the curl of predicted velocity to $0.2^\circ$ in phase ($R = 0.999$), and predicted current the Grad--Shafranov flux to $0.3^\circ$ ($R = 0.990$); the phase-blind control violates both. \emph{Resonance and parity:} the predicted current peak selects the correct rational surface $q{=}m/n$ in 357 of 358 cases from a median of 4--7 candidates, localizing it to 0.67 cells against the truth's own 0.66, and the predicted flux carries the layer's measured parity signature, signs matching the truth in every inferred $m{=}2$ case. \emph{Magnetic topology:} Poincar\'e sections traced through the predicted flux reproduce the truth's topology, including a $(1,1)$ internal kink island chain, at 6.7\% puncture-density total variation against 11.5\% for the conventionally trained twin (Figure~\ref{fig:poincare}; controls in Appendix~\ref{app:poincare}). \emph{Parameter trends:} within unseen equilibria the predicted $\gamma(n)$ selects the fastest-growing harmonic in all 55 cases (chance $\approx 0.5$) at half the twin's spacing error;
and on the two core $n=1$ linear runs on which JOREK is benchmarked~\citep{pamela2020fullmhd}, the surrogate carries the truth's measured trends: the resistivity response in every inferred-$m$ stratum (the twin sits CI-separated low on $m{=}1$) and the layer-width scaling (Figure~\ref{fig:scaling}b).

\paragraph{Transfer to an out-of-distribution resistivity scan.}
\label{sec:transfer}
We test transfer on the controlled resistivity scan of Section~\ref{sec:scan}, an equilibrium family the network never saw. The test asks whether the pretrained model exhibits a growth-rate response to $\eta$, and how many simulations it takes to read that response out.

\paragraph{Reference scan against theory.}
Fitted over the native runs, our reference growth rates give $\log_{10}\gamma$ against $\log_{10}\eta$ slopes of $+0.67$ on the kink branch and $+0.51$ on the tearing branch, against the resistivity scalings $\eta^{1/3}$ and $\eta^{3/5}$; the published model-303 kink points of \citet{pamela2020fullmhd} give $+0.42$ over the same window at the same viscosity. The tearing branch sits close to its asymptote, although its $P_m$ exceeds unity below $\eta \approx 10^{-8}$, outside the inviscid limit of the $\eta^{3/5}$ scaling. The kink branch is steeper than both the analytic exponent and the published points, a gap we cannot close without the original inputs. We therefore treat the reconstruction as an out-of-distribution test case of the same family rather than a reproduction of the published benchmark.

\begin{figure}[tbp]
    \centering
    \includegraphics[width=0.48\linewidth]{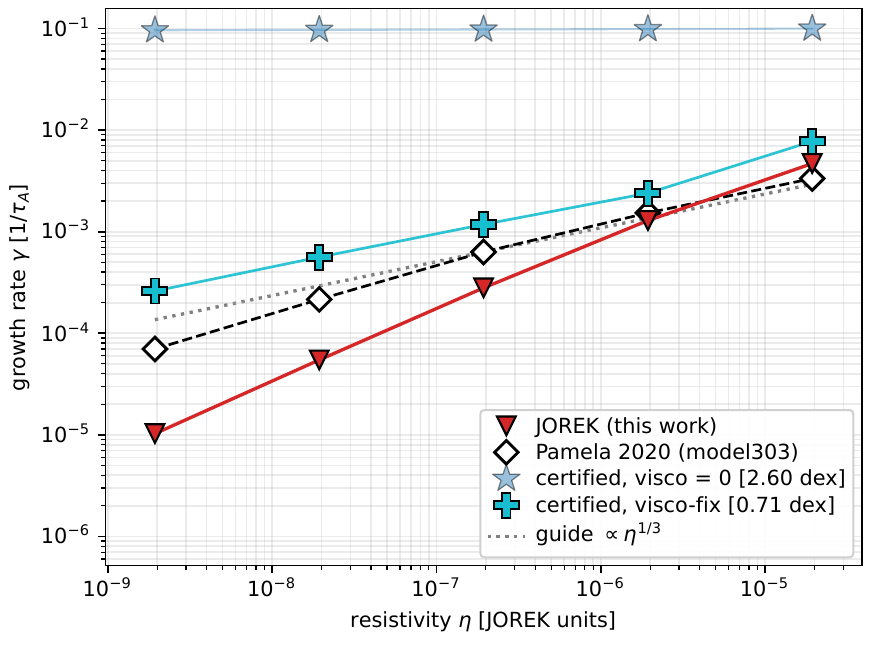}
  \hfill
  \includegraphics[width=0.48\linewidth]{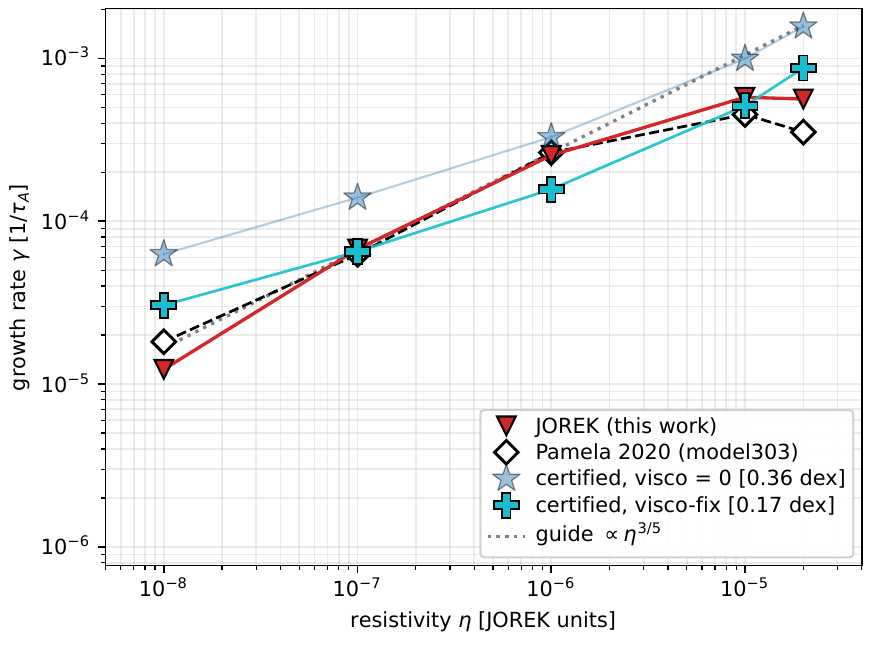}
  \caption{The controlled resistivity scan, ensemble, zero-shot: no fitting of any kind; the pretrained model evaluated directly. Left: $1/1$ internal kink. Right: $2/1$ tearing mode. Red: our JOREK reference runs. White diamonds: the published reduced-MHD points of \citet{pamela2020fullmhd}, plotted for reference. Stars: prediction under raw inputs at the benchmark's own viscosity. Crosses: the same prediction with $\mu$ alone replaced by the training median $3.25\times10^{-6}$. Dotted: the analytic resistive scalings, $\eta^{1/3}$ for the kink and $\eta^{3/5}$ for tearing \citep{coppi1976kink, furth1963tearing}, as guides only. "Certified" in the legends denotes the four-member ensemble.}
  \label{fig:pamela_scan}
\end{figure}

\begin{figure}[tbp]
    \centering
    \includegraphics[width=\scanpanel]{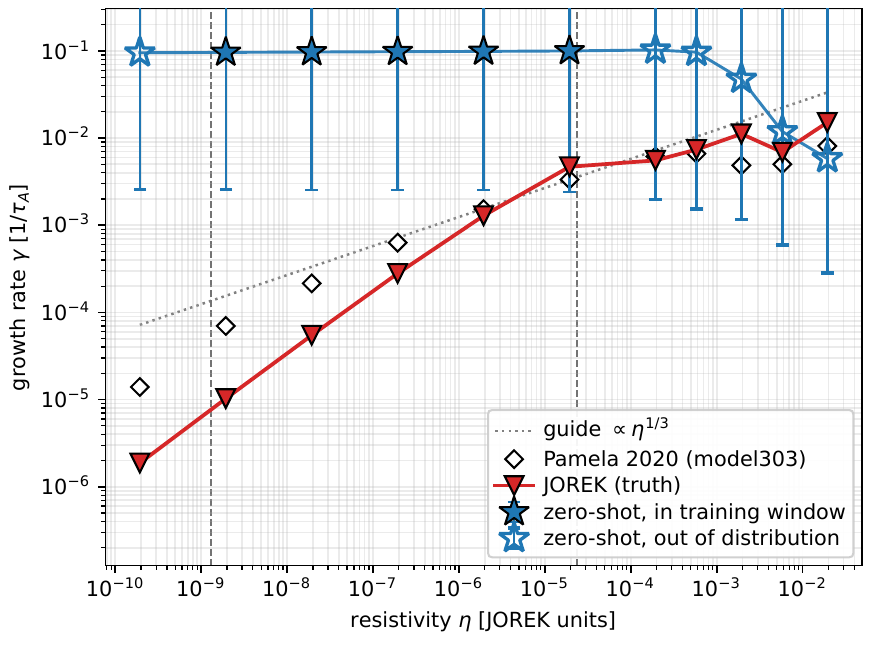}
    \scangap
    \includegraphics[width=\scanpanel]{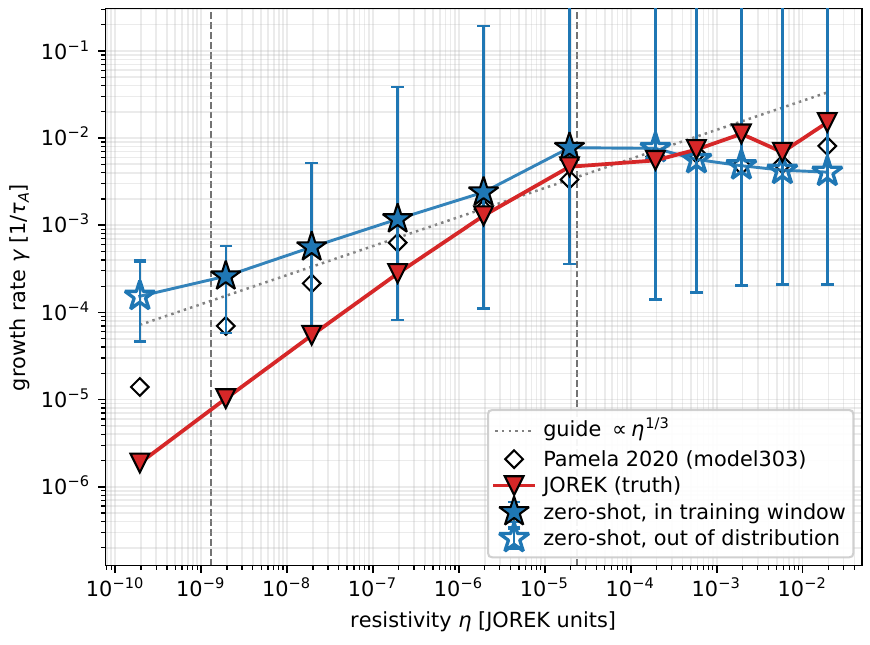}\\[0.6em]
    \includegraphics[width=\scanpanel]{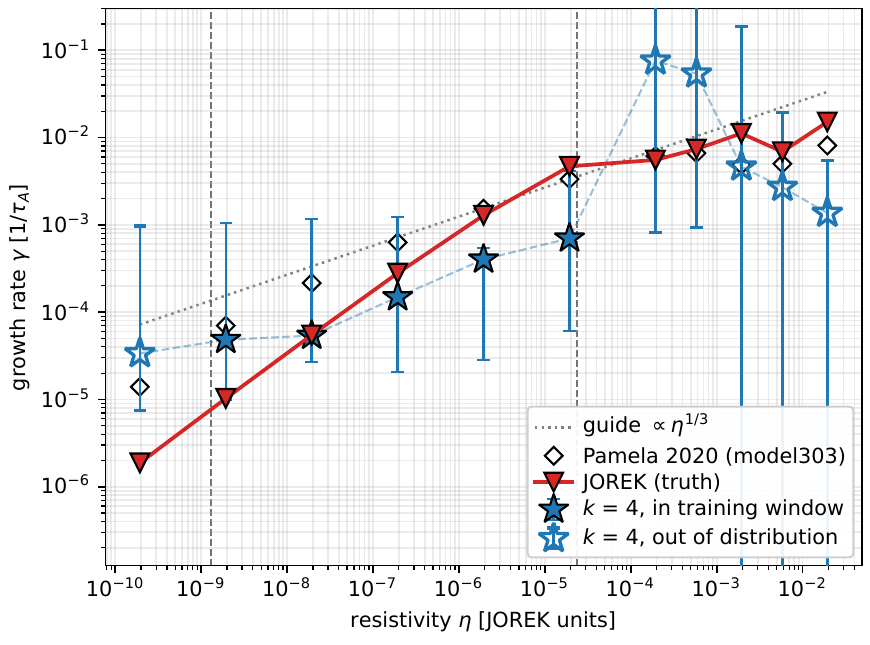}
    \scangap
    \includegraphics[width=\scanpanel]{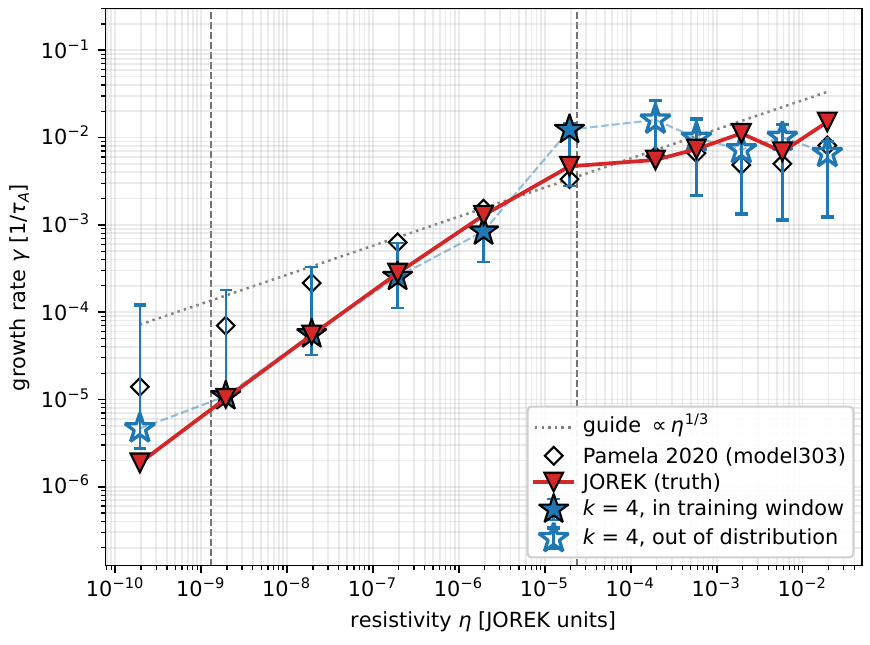}
  \caption{Internal kink, ensemble, before and after the head
  refit. Top: zero-shot, no fitting, every run scored, bars spanning the four members. Bottom: after refitting the 22k-parameter head on $k{=}4$ runs, showing only runs held out of the fit; each star is the median over the 200 nested orderings in which that run was held out, and bars span the 16th to 84th percentile, so the joining line is a guide to the eye rather than one model's curve. Within each row, the left is the raw inputs at $\mu = 0$ and the right is the viscosity-corrected. Filled stars are inside the training $\eta$ window marked by the dashed edges at $1.293\times10^{-9}$ and $2.394\times10^{-5}$; open stars are outside it. Above $\eta \approx 10^{-4}$, the raw-input fit becomes unstable across orderings, with bars exceeding the axis.}
  \label{fig:pamela_fewshot_kink}
\end{figure}

\begin{figure}[tbp]
    \centering
    \includegraphics[width=\scanpanel]{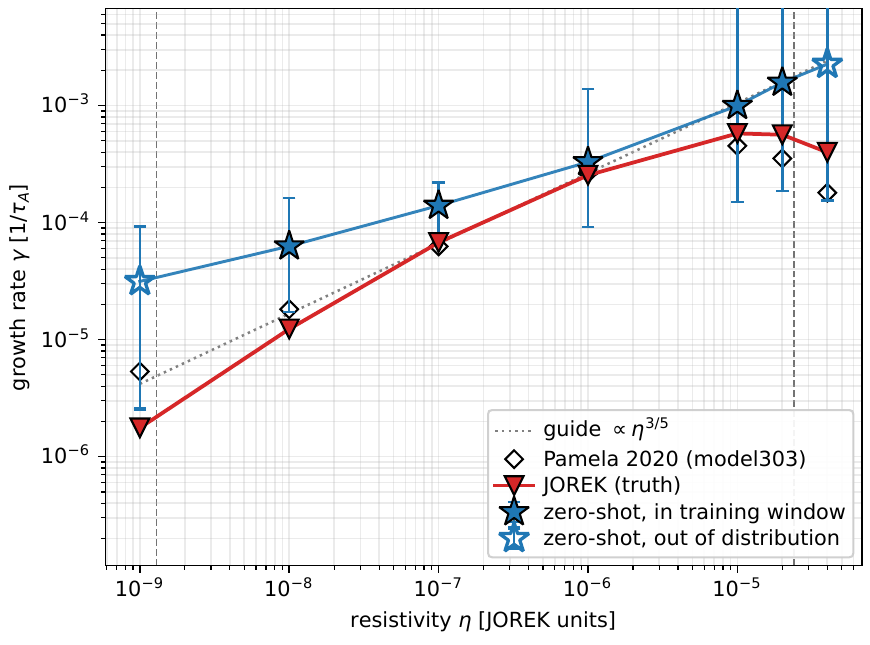}
    \scangap
    \includegraphics[width=\scanpanel]{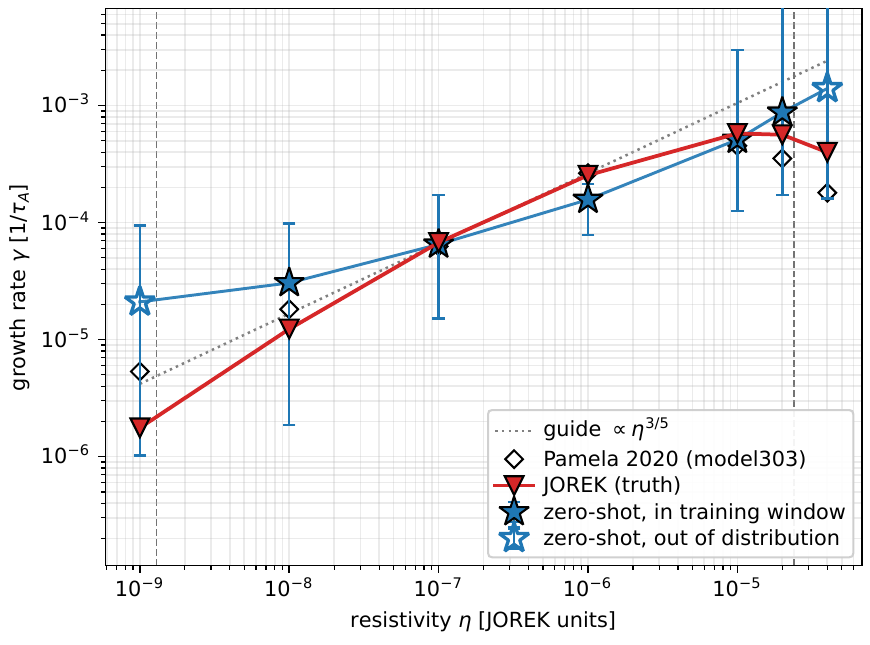}\\[0.6em]
    \includegraphics[width=\scanpanel]{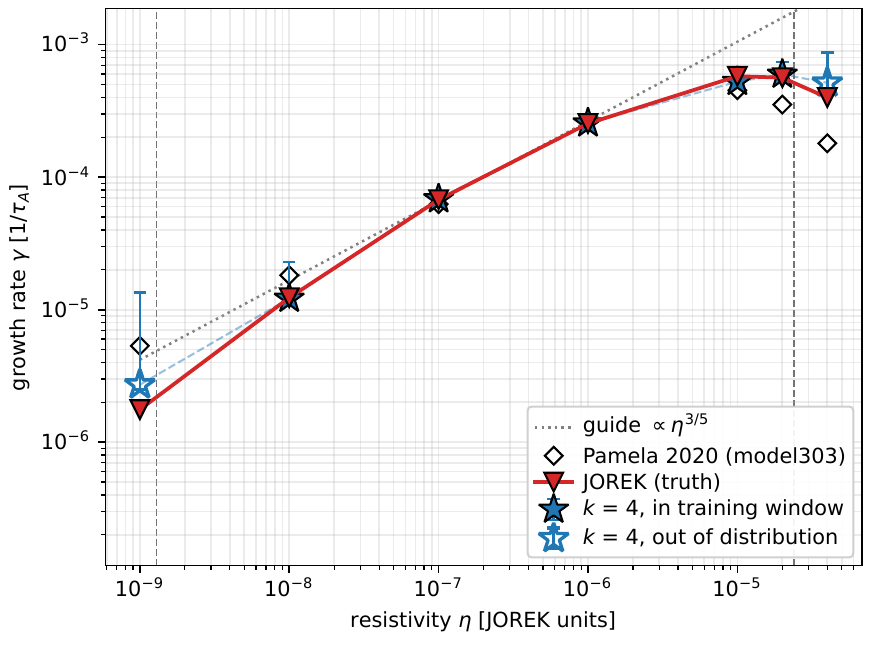}
    \scangap
    \includegraphics[width=\scanpanel]{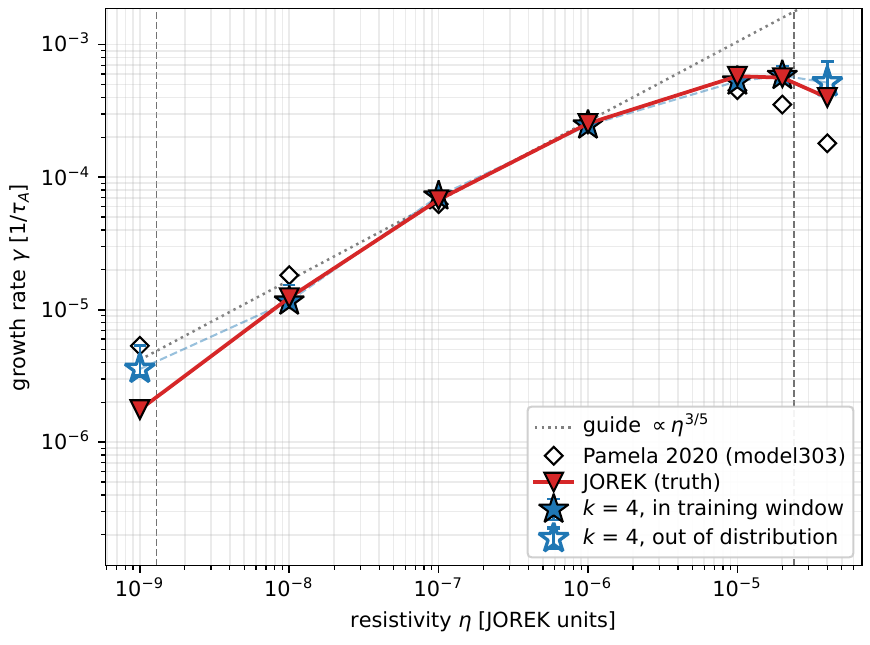}
  \caption{Tearing mode, ensemble, before and after the head
  refit; rows, columns and markers as in Figure~\ref{fig:pamela_fewshot_kink}; raw inputs carry the reference viscosity $\mu = 10^{-8}$. Three held-out runs at $k{=}4$. Native error falls from $0.36$ to $0.03$~dex on raw inputs and from $0.17$ to $0.04$~dex with visco-fix, so on this branch, the refit closes almost all of the gap regardless of which input set it starts from.}
  \label{fig:pamela_fewshot_tear}
\end{figure}

\paragraph{Zero-shot response.}
On the kink branch, repairing the one out-of-range input changes the prediction. Under raw inputs, the model returns essentially the same growth rate at every $\eta$ in the scan: the response is flat, and error over the native runs is $2.60$~dex. With $\mu$ alone replaced, the prediction acquires a slope in $\eta$ close to the reference runs' own, and error falls to $0.71$~dex, still with no fitting of any kind (Figure~\ref{fig:pamela_scan}, which shows both input sets on one axis per branch). The out-of-range $P_m$ suppresses the resistivity response itself, so no offset or rescaling can recover it.
Figures~\ref{fig:pamela_fewshot_kink} and~\ref{fig:pamela_fewshot_tear} show the two branches before and after the refit, distinguishing native from out-of-distribution runs.

\paragraph{Few-shot results.}
Table~\ref{tab:pamela} gives the eight values fixed before scoring. The numbers in brackets below are baselines scored afterward on the same held-out sets and are not in the table: the pretrained model with no refitting, and a straight line in $\log_{10}\eta$ fitted to the same $k$ runs. On the kink branch, the head fit roughly halves the zero-shot error and beats the line ($0.406$~dex), though only in 72\% of the 200 orderings. On the tearing branch, the zero-shot error over the same held-out sets is $0.556$~dex, so three runs cut it threefold, and the head fit beats the line ($0.344$~dex) in every ordering. On native runs, the tearing refit closes almost all of the remaining gap (Figure~\ref{fig:pamela_fewshot_tear}). The refit does not repair an out-of-range input: with $\mu = 0$ left in the kink inputs, the line baseline beats it (rows 7 and 8). The leave-one-out rows measure interpolation inside a dense set of fitted runs rather than transfer, and the $0.010$~dex tearing value lies within a factor of two of the growth rate label noise (Appendix~\ref{app:dataset}).

\begin{table}[tbp]
  \caption{Held-out error on the scan, in dex: the median over 200
  permutations of the per-permutation mean absolute error over held-out runs, at the 300-step head snapshot, against our reference simulations. These are the eight values pre-registered before scoring. The lower block holds out a single run and is not comparable to the block above it. Row 7 is the one combination where the surrogate is beaten by the no-surrogate line baseline of row 8.}
  \label{tab:pamela}
  \centering
  \small
  \begin{tabular}{cllllcc}
    \toprule
    \# & Branch & Inputs & Estimator & $k$ & Held out & Error (dex)\\
    \midrule
    3 & kink $1/1$ & visco-fix & ensemble, zero-shot      & 4  & 7 & 0.622\\
    1 & kink $1/1$ & visco-fix & ensemble $+$ head fit    & 4  & 7 & \textbf{0.323}\\
    4 & tear $2/1$ & raw             & ensemble $+$ head fit    & 3  & 4 & \textbf{0.183}\\
    \midrule
    \multicolumn{7}{l}{\emph{Leave-one-out; consistency checks, not precision measurements}}\\
    2 & kink $1/1$ & visco-fix & ensemble $+$ head fit    & 10 & 1 & 0.077\\
    5 & tear $2/1$ & raw             & ensemble $+$ head fit    & 6  & 1 & 0.020\\
    6 & tear $2/1$ & visco-fix & ensemble $+$ head fit    & 6  & 1 & 0.010\\
    7 & kink $1/1$ & raw ($\mu{=}0$) & ensemble $+$ head fit    & 10 & 1 & 0.509\\
    8 & kink $1/1$ & raw ($\mu{=}0$) & $\log\eta$ line baseline & 10 & 1 & 0.315\\
    \bottomrule
  \end{tabular}
\end{table}

\paragraph{Where the adaptation has to happen.}
Table~\ref{tab:peft} compares ways of adapting the model to the scan, from refitting the head alone to fine-tuning the whole network, on both branches, using the four-member ensemble with visco-fix inputs throughout. At $k{=}4$, refitting the head alone comes within $0.07$~dex of the best kink result (LoRA with a frozen head and full fine-tuning, both $0.27$) and ties the best tearing result ($0.10$), while training $42\times$ and $6{,}700\times$ fewer parameters than LoRA and full fine-tuning. Adapting the trunk with LoRA adapters~\citep{hu2021lora} or full fine-tuning therefore gains little, at most $0.07$~dex on kink and nothing on tearing, which is the expected outcome if the trunk's representation of the resistivity response is already adequate and the error lies in the map from that representation to a growth rate. The two branches agree on this, which one branch alone could not establish. Four runs may also be too few to constrain a trunk update, so the comparison bounds what adaptation buys at this data budget rather than proving the trunk exact. At $k = 1$, adaptation mostly hurts: on tear, every method is worse than the pretrained model, and on kink, every method except LoRA with a frozen head and full fine-tuning is. One run does not constrain the head's map, and the fit overshoots it. The gain appears by three to four runs, which is where we quote the scan. Error continues to fall at larger $k$ (Table~\ref{tab:pamela}), but the held-out set shrinks as $k$ grows, so those points measure interpolation between an increasingly dense set of fitted runs rather than further transfer.

\begin{table}[tbp]
  \caption{Adaptation methods on both branches for ensemble: held-out
  MAE in dex. Both column groups use visco-fix inputs. The two groups are separate few-shot problems on different error scales and are never averaged. Accuracy is the share of held-out runs predicted within a factor of two in $\gamma$ at $k{=}4$. BitFit trains the additive bias terms of the trunk and the head; the last-block row trains the final trunk block and the head; and LoRA inserts rank-4 adapters alongside the trunk's linear layers. These rows use eight shot-orderings rather than the 200 permutations of Table~\ref{tab:pamela}, and full fine-tuning uses the first four of those eight, so levels compare within this table only. Snapshots are 300 steps for the head, 200 for the other parameter-efficient methods, and 100 for full fine-tuning.}
  \label{tab:peft}
  \centering
  \small
  \resizebox{\linewidth}{!}{%
  \begin{tabular}{lrcccccccc}
    \toprule
    & & \multicolumn{4}{c}{Internal kink $1/1$ (11 runs)} & \multicolumn{4}{c}{Tearing $2/1$ (7 runs)}\\
    \cmidrule(lr){3-6}\cmidrule(lr){7-10}
    Method & Trainable & $k{=}1$ & $k{=}2$ & $k{=}4$ & Acc.\ $\le 2\times$ & $k{=}1$ & $k{=}2$ & $k{=}4$ & Acc.\ $\le 2\times$\\
    \midrule
    Pretrained, no fitting    & 0               & 0.66 & 0.62 & 0.66          & 43\% & 0.36 & 0.31 & 0.43          & 54\%\\
    Head only                 & 21{,}768        & 0.92 & 0.62 & 0.34          & 50\% & 0.53 & 0.52 & \textbf{0.10} & 83\%\\
    BitFit                    & 93{,}800        & 0.82 & 0.62 & 0.43          & 45\% & 0.47 & 0.57 & 0.16          & 75\%\\
    Last trunk block          & 163{,}880       & 1.22 & 0.90 & 0.38          & 57\% & 0.88 & 0.51 & 0.28          & 75\%\\
    LoRA $r{=}4$, frozen head & 914{,}716       & 0.61 & 0.67 & \textbf{0.27} & 66\% & 0.45 & 0.60 & 0.15          & 79\%\\
    LoRA $r{=}4$              & 936{,}484       & 0.83 & 0.68 & 0.32          & 52\% & 0.47 & 0.56 & \textbf{0.10} & 83\%\\
    Full fine-tune            & 144{,}871{,}926 & 0.57 & 0.67 & \textbf{0.27} & 64\% & 0.49 & 0.60 & 0.14          & 83\%\\
    \bottomrule
  \end{tabular}}
\end{table}

\FloatBarrier
\section{Conclusion and Outlook}
\label{sec:limitations}
\label{sec:outlook}
\modelname\ is a first step toward a new way of designing and operating fusion reactors: a learned model that returns the full eigenmode and growth rate for every configuration it is shown. Used as a screen, the model already changes how simulation is spent: at the conservative operating point it passes no unstable configuration, clears the stable majority in 0.05~s per input on one GPU (2.3~s on a single CPU core) from the design parameters, against 1.3--4.3~h on 32--64 cores (83--137 core-hours) for a JOREK linear run, up to $10^6\times$ faster for all four mode numbers of a design (Appendix~\ref{app:e2e}), and refers roughly 7\% of borderline cases to a full solve (Appendix~\ref{app:systems}). A design study can then escalate only the uncertain equilibria to JOREK, closing an active-learning loop.{} The same workflow reaches past the training distribution. On a resistivity scan of a circular equilibrium held at fixed geometry, refitting only the 22k-parameter growth rate head on four simulations halves the held-out internal kink error, three cut the tearing error threefold, and fine-tuning the whole network gains at most $0.07$~dex more. Porting to a new equilibrium family may then cost runs in the single digits rather than a campaign, under two conditions the scan makes explicit: inputs that leave the training range, such as $\mu = 0$, must be repaired first, and the evidence so far covers growth rates on two branches of one geometry, not the eigenmode fields.{}

\paragraph{Limitations.}
  These experiments show that \modelname\ reproduces JOREK's linear stability calculation within the simulated distribution: one code, one device-geometry family, linear reduced-MHD physics in the plasma core with no X-point, no edge or scrape-off layer (where peeling--ballooning modes govern a real tokamak), no nonlinear phase~\citep{paischer2025gyroswin, carey2025fno}, and no experimental data. Outside that distribution, the evidence consists of one external resistivity scan, in which growth rates transfer (Section~\ref{sec:transfer}). Accuracy also falls with distance from the training set, most clearly in
  the test split's farthest quartile (Appendix~\ref{app:results}).
  The remaining resistive-layer error may be a resolution limit of the $128^2$ grid; re-simulating a slice at finer resolution around the rational surface would test this.

  \paragraph{Outlook.}
  Because the model is differentiable from equilibrium to full field, sparse, indirect diagnostics could, in principle, be assimilated through it. Trained jointly on simulation and experiments, a successor could hold a Pareto point that no solver can, closer to the measured ground truth at orders-of-magnitude lower cost. Run in real time, such a model could form part of a digital twin that reads a device's sparse signals and flags approaching instability, which would first need edge physics, nonlinear evolution, and validation against experiment. We leave these steps, aimed at designing and optimizing the next generation of devices, to future work.

\FloatBarrier
\bibliographystyle{unsrtnat}
\bibliography{refs}

\begin{thebibliography}{25}
\providecommand{\natexlab}[1]{#1}
\providecommand{\url}[1]{\texttt{#1}}
\expandafter\ifx\csname urlstyle\endcsname\relax
  \providecommand{\doi}[1]{doi: #1}\else
  \providecommand{\doi}{doi: \begingroup \urlstyle{rm}\Url}\fi

\bibitem[Zhu et~al.(2023)Zhu, Yu, Chen, Shi, Ge, Wang, Luan, Sun, and
  Wang]{zhu2023fishbone}
X.~L. Zhu, L.~M. Yu, W.~Chen, P.~W. Shi, W.~L. Ge, F.~Wang, Q.~B. Luan, H.~E.
  Sun, and Z.~X. Wang.
\newblock Hybrid-kinetic simulation of synergy between fishbone/sawtooth and
  tearing mode-induced energetic-ion transport in a tokamak plasma.
\newblock \emph{Nuclear Fusion}, 63:\penalty0 036014, 2023.
\newblock \doi{10.1088/1741-4326/acb221}.

\bibitem[Hoelzl et~al.(2021)Hoelzl, Huijsmans, Pamela, B{\'e}coulet, Nardon,
  Artola, Nkonga, et~al.]{hoelzl2021jorek}
M.~Hoelzl, G.~T.~A. Huijsmans, S.~J.~P. Pamela, M.~B{\'e}coulet, E.~Nardon,
  F.~J. Artola, B.~Nkonga, et~al.
\newblock The {JOREK} non-linear extended {MHD} code and applications to
  large-scale instabilities and their control in magnetically confined fusion
  plasmas.
\newblock \emph{Nuclear Fusion}, 61\penalty0 (6):\penalty0 065001, 2021.
\newblock \doi{10.1088/1741-4326/abf99f}.

\bibitem[Pamela et~al.(2020)Pamela, Bhole, Huijsmans, Nkonga, Hoelzl, Krebs,
  Strumberger, and {JET Contributors}]{pamela2020fullmhd}
S.~J.~P. Pamela, A.~Bhole, G.~T.~A. Huijsmans, B.~Nkonga, M.~Hoelzl, I.~Krebs,
  E.~Strumberger, and {JET Contributors}.
\newblock Extended full-{MHD} simulation of non-linear instabilities in tokamak
  plasmas.
\newblock \emph{Physics of Plasmas}, 27:\penalty0 102510, 2020.
\newblock \doi{10.1063/5.0018208}.

\bibitem[Artola et~al.(2021)Artola, Sovinec, Jardin, Hoelzl, Krebs, and
  Clauser]{artola2021vde}
F.~J. Artola, C.~R. Sovinec, S.~C. Jardin, M.~Hoelzl, I.~Krebs, and C.~Clauser.
\newblock 3{D} simulations of vertical displacement events in tokamaks: A
  benchmark of {M3D-C1}, {NIMROD}, and {JOREK}.
\newblock \emph{Physics of Plasmas}, 28:\penalty0 052511, 2021.
\newblock \doi{10.1063/5.0037115}.

\bibitem[Haverkort et~al.(2016)Haverkort, de~Blank, Huysmans, Pratt, and
  Koren]{haverkort2016fmhd}
J.~W. Haverkort, H.~J. de~Blank, G.~T.~A. Huysmans, J.~Pratt, and B.~Koren.
\newblock Implementation of the full viscoresistive magnetohydrodynamic
  equations in a nonlinear finite element code.
\newblock \emph{Journal of Computational Physics}, 316:\penalty0 281--302,
  2016.
\newblock \doi{10.1016/j.jcp.2016.04.007}.

\bibitem[Piccione et~al.(2020)Piccione, Berkery, Sabbagh, and
  Andreopoulos]{piccione2020pb}
A.~Piccione, J.~W. Berkery, S.~A. Sabbagh, and Y.~Andreopoulos.
\newblock Physics-guided machine learning approaches to predict the ideal
  stability properties of fusion plasmas.
\newblock \emph{Nuclear Fusion}, 60:\penalty0 046033, 2020.
\newblock \doi{10.1088/1741-4326/ab7597}.

\bibitem[Bruncrona et~al.(2025)Bruncrona, Kit, J{\"a}rvinen, Saarelma,
  Frassinetti, and {JET Contributors}]{bruncrona2025pb}
A.~M. Bruncrona, A.~Kit, A.~E. J{\"a}rvinen, S.~Saarelma, L.~Frassinetti, and
  {JET Contributors}.
\newblock Machine learning surrogate model for ideal peeling--ballooning
  pedestal {MHD} stability.
\newblock \emph{Physics of Plasmas}, 32:\penalty0 092501, 2025.
\newblock \doi{10.1063/5.0282085}.

\bibitem[Xu et~al.(2024)Xu, Luan, Li, Chen, and Wang]{xu2024dtm}
J.~Xu, Q.~Luan, H.~Li, Y.~Chen, and Z.~Wang.
\newblock Neural network based fast prediction of double tearing modes in
  advanced tokamak plasmas.
\newblock \emph{Physics of Plasmas}, 31:\penalty0 122113, 2024.
\newblock \doi{10.1063/5.0229910}.

\bibitem[Liu et~al.(2024)Liu, Qiu, Fu, Xiao, Chen, Wang, and
  Wei]{liu2024fishbone}
Z.~Y. Liu, H.~R. Qiu, G.~Y. Fu, Y.~Xiao, Y.~C. Chen, Z.~J. Wang, and Y.~X. Wei.
\newblock Prediction of fishbone linear instability in tokamaks with machine
  learning methods, 2024.
\newblock arXiv:2402.15051.

\bibitem[Kumar et~al.(2026)Kumar, Clauser, Golfinopoulos, Rea, Capersene,
  Boyer, {SPARC Team}, and {Alcator C-Mod Team}]{kumar2026vertical}
A.~Kumar, C.~Clauser, T.~Golfinopoulos, C.~Rea, F.~Capersene, D.~Boyer, {SPARC
  Team}, and {Alcator C-Mod Team}.
\newblock Physics attention transformer surrogate for rapid vertical
  instability growth rate prediction: {Alcator C-Mod} to {SPARC}, 2026.
\newblock arXiv:2608.24785.

\bibitem[Carey et~al.(2025)Carey, Zanisi, Pamela, Gopakumar, Omotani, Buchanan,
  Brandstetter, Paischer, Galletti, and Setinek]{carey2025fno}
N.~Carey, L.~Zanisi, S.~Pamela, V.~Gopakumar, J.~Omotani, J.~Buchanan,
  J.~Brandstetter, F.~Paischer, G.~Galletti, and P.~Setinek.
\newblock Neural operator surrogate models of plasma edge simulations:
  feasibility and data efficiency.
\newblock \emph{Nuclear Fusion}, 65\penalty0 (10):\penalty0 106010, 2025.
\newblock \doi{10.1088/1741-4326/adfdfb}.
\newblock arXiv:2502.17386.

\bibitem[Paischer et~al.(2025)Paischer, Galletti, Hornsby, Setinek, Zanisi,
  Carey, Pamela, and Brandstetter]{paischer2025gyroswin}
F.~Paischer, G.~Galletti, W.~Hornsby, P.~Setinek, L.~Zanisi, N.~Carey,
  S.~Pamela, and J.~Brandstetter.
\newblock {GyroSwin}: 5{D} surrogates for gyrokinetic plasma turbulence
  simulations.
\newblock In \emph{Advances in Neural Information Processing Systems
  (NeurIPS)}, 2025.
\newblock arXiv:2510.07314.

\bibitem[Gopakumar et~al.(2023)Gopakumar, Pamela, Zanisi, Li, Anandkumar, and
  {MAST Team}]{gopakumar2023fno}
V.~Gopakumar, S.~Pamela, L.~Zanisi, Z.~Li, A.~Anandkumar, and {MAST Team}.
\newblock Fourier neural operator for plasma modelling, 2023.
\newblock arXiv:2302.06542.

\bibitem[Chang et~al.(2025)Chang, Gao, Yin, and Zhao]{chang2025dltm}
Z.~Chang, B.~Gao, R.~Yin, and X.~Zhao.
\newblock {DLTM}: a deep learning method for tearing mode simulation and
  prediction.
\newblock \emph{Nuclear Fusion}, 65:\penalty0 086009, 2025.
\newblock \doi{10.1088/1741-4326/ade891}.

\bibitem[Furth et~al.(1963)Furth, Killeen, and Rosenbluth]{furth1963tearing}
H.~P. Furth, J.~Killeen, and M.~N. Rosenbluth.
\newblock Finite-resistivity instabilities of a sheet pinch.
\newblock \emph{Physics of Fluids}, 6\penalty0 (4):\penalty0 459--484, 1963.
\newblock \doi{10.1063/1.1706761}.

\bibitem[Coppi et~al.(1976)Coppi, Galvao, Pellat, Rosenbluth, and
  Rutherford]{coppi1976kink}
B.~Coppi, Ricardo Galvao, R.~Pellat, M.~Rosenbluth, and P.~Rutherford.
\newblock Resistive internal kink modes.
\newblock \emph{Soviet Journal of Plasma Physics}, 07 1976.

\bibitem[Ronneberger et~al.(2015)Ronneberger, Fischer, and
  Brox]{ronneberger2015unet}
O.~Ronneberger, P.~Fischer, and T.~Brox.
\newblock {U-Net}: Convolutional networks for biomedical image segmentation.
\newblock In \emph{Medical Image Computing and Computer-Assisted Intervention
  (MICCAI)}, pages 234--241, 2015.
\newblock \doi{10.1007/978-3-319-24574-4_28}.

\bibitem[Li et~al.(2021)Li, Kovachki, Azizzadenesheli, Liu, Bhattacharya,
  Stuart, and Anandkumar]{li2021fno}
Z.~Li, N.~Kovachki, K.~Azizzadenesheli, B.~Liu, K.~Bhattacharya, A.~Stuart, and
  A.~Anandkumar.
\newblock Fourier neural operator for parametric partial differential
  equations.
\newblock In \emph{International Conference on Learning Representations
  (ICLR)}, 2021.

\bibitem[Dosovitskiy et~al.(2021)Dosovitskiy, Beyer, Kolesnikov, Weissenborn,
  Zhai, Unterthiner, Dehghani, Minderer, Heigold, Gelly, Uszkoreit, and
  Houlsby]{dosovitskiy2021vit}
A.~Dosovitskiy, L.~Beyer, A.~Kolesnikov, D.~Weissenborn, X.~Zhai,
  T.~Unterthiner, M.~Dehghani, M.~Minderer, G.~Heigold, S.~Gelly, J.~Uszkoreit,
  and N.~Houlsby.
\newblock An image is worth 16x16 words: Transformers for image recognition at
  scale.
\newblock In \emph{International Conference on Learning Representations
  (ICLR)}, 2021.

\bibitem[Alkin et~al.(2024)Alkin, F{\"u}rst, Schmid, Gruber, Holzleitner, and
  Brandstetter]{alkin2024upt}
B.~Alkin, A.~F{\"u}rst, S.~Schmid, L.~Gruber, M.~Holzleitner, and
  J.~Brandstetter.
\newblock Universal physics transformers: A framework for efficiently scaling
  neural operators.
\newblock In \emph{Advances in Neural Information Processing Systems
  (NeurIPS)}, 2024.

\bibitem[Hagnberger et~al.(2025)]{hagnberger2025calmpde}
J.~Hagnberger et~al.
\newblock {CALM-PDE}: Continuous and adaptive convolutions for latent space
  modeling of time-dependent {PDEs}.
\newblock In \emph{Advances in Neural Information Processing Systems
  (NeurIPS)}, 2025.
\newblock arXiv:2505.12944.

\bibitem[Lakshminarayanan et~al.(2017)Lakshminarayanan, Pritzel, and
  Blundell]{lakshminarayanan2017ensembles}
B.~Lakshminarayanan, A.~Pritzel, and C.~Blundell.
\newblock Simple and scalable predictive uncertainty estimation using deep
  ensembles.
\newblock In \emph{Advances in Neural Information Processing Systems},
  volume~30, 2017.
\newblock \doi{10.48550/arXiv.1612.01474}.

\bibitem[Hu et~al.(2022)Hu, Shen, Wallis, Allen-Zhu, Li, Wang, Wang, and
  Chen]{hu2021lora}
E.~J. Hu, Y.~Shen, P.~Wallis, Z.~Allen-Zhu, Y.~Li, S.~Wang, L.~Wang, and
  W.~Chen.
\newblock {LoRA}: Low-rank adaptation of large language models.
\newblock In \emph{International Conference on Learning Representations
  (ICLR)}, 2022.

\bibitem[Lei et~al.(2018)Lei, G'Sell, Rinaldo, Tibshirani, and
  Wasserman]{lei2018conformal}
J.~Lei, M.~G'Sell, A.~Rinaldo, R.~J. Tibshirani, and L.~Wasserman.
\newblock Distribution-free predictive inference for regression.
\newblock \emph{Journal of the American Statistical Association}, 113\penalty0
  (523):\penalty0 1094--1111, 2018.
\newblock \doi{10.1080/01621459.2017.1307116}.

\bibitem[Hastie et~al.(1987)Hastie, Hender, Carreras, Charlton, and
  Holmes]{hastie1987kinkscaling}
R.~J. Hastie, T.~C. Hender, B.~A. Carreras, L.~A. Charlton, and J.~A. Holmes.
\newblock Stability of ideal and resistive internal kink modes in toroidal
  geometry.
\newblock \emph{Physics of Fluids}, 30\penalty0 (6):\penalty0 1756--1766, 1987.
\newblock \doi{10.1063/1.866242}.

\end{thebibliography}

\newpage
\appendix

\section{Tokamak geometry and Symbols}
\label{app:geometry&symbols}

\begin{figure}[h]
    \centering
    \includegraphics[width=\linewidth]{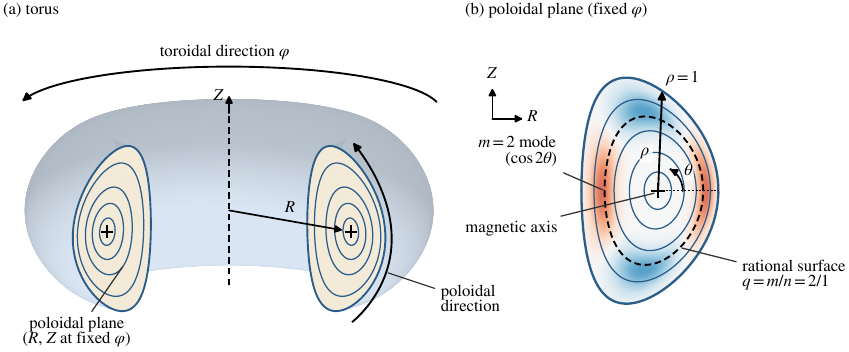}
    \caption{Schematic of a tokamak. (a) The plasma fills a torus; $\varphi$ runs the long way around it, and a vertical cut at fixed $\varphi$ gives the poloidal plane. (b) The poloidal plane: nested flux surfaces around the magnetic axis, labeled by the normalized radial coordinate $\rho$, with poloidal angle $\theta$. The dashed line is the rational surface $q = 2$, where an $(m,n) = (2,1)$ mode resonates; color shows the mode's $\cos 2\theta$ pattern.}
    \label{fig:geometry}
  \end{figure}

  A tokamak confines the plasma in a torus (Figure~\ref{fig:geometry}a). Positions use cylindrical coordinates $(R, Z, \varphi)$: $R$ is the distance from the torus's vertical axis, $Z$ the height, and $\varphi$ the angle the long way around the ring, the \emph{toroidal} direction. A vertical cut at fixed $\varphi$ is the \emph{poloidal plane} (Figure~\ref{fig:geometry}b); the short way around the plasma's cross-section is the \emph{poloidal} direction, with angle $\theta$.

  In equilibrium, magnetic field lines wind helically around the torus and lie on nested surfaces of constant poloidal flux $\Psi$, the flux surfaces, which shrink to a single line at the center, the magnetic axis. The normalized radial coordinate $\rho$ labels the surfaces, from 0 at the axis to 1 at the edge. The safety factor $q$ is the number of toroidal turns a field line makes per poloidal turn on a given surface.

  The equilibrium is axisymmetric: it is the same on every poloidal plane, so it is a 2D field on $(R, Z)$. A perturbation therefore splits into independent toroidal harmonics $e^{in\varphi}$, where the toroidal mode number $n$ counts the perturbation's periods around the torus. Around the poloidal direction, the mode has a dominant poloidal number $m$. A mode $(m, n)$ resonates on the rational surface $q = m/n$, where its helical pitch matches that of the field lines; tearing modes and the resistive internal kink grow there. All fields in this paper, inputs and outputs, are 2D maps on the poloidal plane; the $\varphi$ dependence is the known factor $e^{in\varphi}$ (Equation~\eqref{eq:ansatz}).

\begin{table}[h]
  \caption{Symbols used throughout the paper. The seven perturbation fields
  are the channels of the predicted complex mode shape $\hat{f}_n$.}
  \label{tab:glossary}
  \centering
  \small
  \begin{tabular}{lp{0.72\linewidth}}
    \toprule
    $\Psi$ & poloidal magnetic flux (perturbation channel)\\
    $u$ & velocity stream function: the potential of the poloidal flow\\
    $J_\phi$ & toroidal current density\\
    $\omega$ & vorticity of the poloidal flow (Laplacian of $u$)\\
    $\rho$ & mass density\\
    $T$ & temperature\\
    $v_\parallel$ & velocity component parallel to the magnetic field\\
    \midrule
    $R, Z, \varphi$ & major-radius, vertical, and toroidal coordinates\\
    $\rho$ (coordinate) & normalized radial coordinate labeling the nested flux surfaces: 0 at the magnetic axis, 1 at the domain's bounding flux surface; the grid's radial axis (128 cells, one cell $=1/128$). Distinct from the density channel $\rho$ above, per standard usage; context disambiguates\\
    $\theta$ & poloidal angle\\
    $n$ & toroidal mode number of the perturbation ($e^{in\varphi}$)\\
    $m$ & poloidal mode number\\
    $q$ & safety factor: field-line pitch per flux surface; a mode with numbers $(m, n)$ resonates at the rational surface $q = m/n$\\
    $\hat{f}_n$ & complex mode shape: the seven-channel 2D eigenmode, defined up to one global phase\\
    $\gamma$ & linear growth rate of the mode, in $\mathrm{s^{-1}}$; predicted as $\log_{10}\gamma$\\
    $\eta$ & resistivity \\
    $\mu$ & kinematic viscosity; enters the network through $P_m$ rather than directly\\
    $P_m$ & magnetic Prandtl number $\mu/\eta$: ratio of viscous to resistive diffusion\\
    $S$ & Lundquist number: ratio of resistive diffusion to Alfv\'en time ($S \propto \frac{1}{\eta} $)\\
    $\beta_0$ & Normalized pressure\\
    $\tau_A$ & Alfv\'en time: transit time of an Alfv\'en wave across the plasma\\
    $\Delta'$ & tearing stability index: jump in the logarithmic derivative of the outer-region perturbed flux across the rational surface; the tearing mode is unstable for $\Delta' > 0$\\
    $\delta$ & resistive-layer width: radial thickness of the sheet at the rational surface where the mode's sharp structure lives (Appendix~\ref{app:results})\\
    $N$ & number of training runs in a data-efficiency cell\\
    $k$ & number of runs the growth rate head is refit on in the few-shot protocol (Section~\ref{sec:transfer})\\
    \bottomrule
  \end{tabular}
\end{table}

\FloatBarrier
\section{Error analysis and model behavior}
\label{app:results}

\paragraph{Nearest-neighbour lookups.}
The $k{=}1$ lookup scores 1.159, worse than predicting the dataset mean (0.921), because a stored field carries full amplitude at the wrong phase and position, while the mean is attenuated toward zero; averaging five neighbors (0.892) recovers most of that attenuation benefit without gaining structure. Relative $L_2$ near 1 thus marks the no-information regime for this task.

\paragraph{Ensembling variants.}
A fifth seed added to the two-stage fine-tune family is a null result (0.3040 vs.\ 0.3027), while the four-seed phase-coherent complex average over the single-stage composite family, the frozen configuration, is CI-separated from zero (paired $-0.019$). Alignment makes the members compatible: members are equivalent up to the mode's global phase, so averaging without alignment cancels amplitude rather than error, while after alignment, the average smooths the sub-cell layer placement on which members disagree most.

\paragraph{Interval construction.}
The growth rate intervals are split-conformal~\citep{lei2018conformal}: a held-out calibration split fixes the $\pm 0.135$ band in $\log_{10}\gamma$ at nominal 90\%, reported at its empirical 91.0\% coverage on the once-run test split in Section~\ref{sec:results}. All eigenmode and $\gamma$ error bands elsewhere are paired case-level bootstrap 95\% CIs on medians; training variability is reported as seed spread over initializations.

\paragraph{Segmented runs.}
A small number of training runs resumed from their own checkpoints after compute interruptions (weights carried over, fresh optimizer state); no run mixes weights across recipes or families.

\paragraph{Ensemble disagreement (conditional-spread proxy).}
As one indication of how far the inputs determine the output, we measure where diversely trained models agree in their errors: six differently conditioned, well-trained models (four feature sets, an attention variant, and a heterogeneous member) are phase-aligned per channel to a common reference, the pixelwise standard deviation across them is taken per case, its $L_2$ norm is divided by the true field's norm, and the median over validation cases is reported. Per channel the proxy reads 0.085 / 0.104 / 0.175 / 0.254 / 0.207 / 0.126 / 0.234 for $\Psi$ / $u$ / $J_\phi$ / $\omega$ / $\rho$ / $T$ / $v_\parallel$: the ensemble of Table~\ref{tab:channels} sits below it on the three smooth channels and above it on the four layer-dominated ones (after oracle sub-cell re-alignment, $\omega$ moves below it as well). The estimate is indicative rather than a hard bound: its per-case calibration against realized error varies by channel (Spearman 0.85 for $\Psi$ down to 0.15 for $\omega$), and it is a statement about model families rather than about the simulator. Pipeline-induced shape noise is bounded separately and is negligible: the mode shape extracted independently from the two stored time frames of each run agrees with per-channel correlation $\approx 1$. A replicate campaign (identical configurations, fresh perturbation seeds) would yield a measured reproducibility floor.

\begin{figure}[t]
  \centering
  \includegraphics[width=\linewidth]{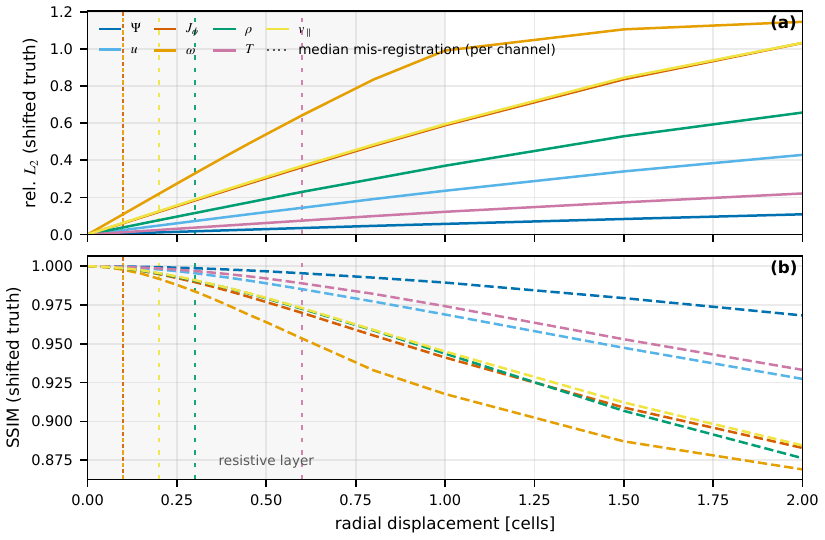}
  \caption{Metric response to pure radial displacement: the \emph{true}
  fields, shifted by $x$ cells (exact Fourier fractional shifts, severity averaged over both directions) and scored against themselves --- no model is involved. (a) Relative $L_2$ rises steeply within half a cell on the layer channels and saturates beyond one cell, which is why integer-cell probes register nothing; (b) SSIM stays above 0.92 at a full cell on every channel, nearly blind to placement. Dashed vertical markers: the median sub-cell mis-registration of \modelname's predictions, measured by the oracle probe; the curve height at a marker is approximately the displacement-attributable share of that channel's error.}
  \label{fig:shiftscan}
\end{figure}

\paragraph{Layer resolution and the Lundquist number.}
The residual concentrates at the rational surface: the four-layer channels place 71--99\% of their error mass within three-layer widths of it, against a 40\% median band area. Per-case error nonetheless correlates only weakly with the Lundquist number (Spearman $-0.02$ to $-0.22$ across channels), and a check for aliasing in the fractional part of $\delta/h$ is null: sharpness sets where the error lives, not how large it is.
In the layer model ($\delta \propto S^{-1/3}$), 7.8\% of validation cases have a layer width below one cell of the $128^2$ grid and 30.2\% below two, the scale to which the metric is most sensitive (Figure~\ref{fig:shiftscan}; a one-cell shift of the \emph{true} $\omega$ alone scores 0.99). An oracle re-alignment decomposes the error by scale: at integer-cell resolution, it recovers almost nothing, but at 0.1-cell resolution, about 70\% of cases take a strictly sub-cell shift, and the oracle removes roughly a fifth of the resistive-channel error ($\omega$ $0.268 \to 0.207$, below its disagreement proxy; $\Psi$ unchanged). The mis-registration is therefore real but sub-cell; the remainder is the sheet's own structure (peak amplitude, width, and splitting). Our working hypothesis is a representation-limited target rather than a data or capacity limit: for the 30\% of cases with $\delta < 2$ cells, the layer sits at the grid scale, where the stored truth changes at the pixel level under sub-cell displacement and dataset growth does not move the ceiling. A slice re-simulated with the mesh packed around the rational surfaces would test this directly and is left for future work; the stored $128^2$ export matches the resolution the simulations were configured to, so the test requires new simulations, not a re-export.

\paragraph{Interpolation and extrapolation.}
Cases are split by nearest-neighbor distance to the training set across the 21 scan parameters, with z-scores based on training statistics. On validation, the quartile medians, nearest to farthest, are 0.260 / 0.287 / 0.278 / 0.304 (CIs overlap); on the once-run test split, read descriptively from stored predictions, the farthest quartile is clearly hardest (0.318 / 0.264 / 0.269 / 0.400) and $\gamma$ RMSE rises from 0.058 to 0.118 across quartiles.

\FloatBarrier
\section{Physics results}
\label{app:physics}

Throughout this appendix, the poloidal number $m$ is a measured label, not a controlled input: each simulation sets $n$ and the geometry, and the dominant $m$ emerges from the self-consistent $q$-profile. The campaigns were designed to excite internal kink and tearing modes, but no family tag is stored per run. Strata assign each row the $m$ of the $q{=}m/n$ crossing nearest the truth's $|J_\phi|$ peak, selected among a median of 4--7 candidate surfaces, and are written inferred $m{=}1$ / $m{=}2$. The claims below concern fidelity within these strata. Accordingly, mode identification in the main text denotes recovery of $n$, of the resonant surface $q{=}m/n$ (357 of 358 validation cases) and of the parity contrast between strata, rather than assignment to a named instability family. The dominant Fourier harmonic of the predicted flux, taken in the grid's geometric poloidal angle at the radius nearest a rational surface, matches the truth in 82.7\% of cases (93.9\% within $\pm 1$; always predicting the most common harmonic scores 45\%). This geometric harmonic is not the inferred $m$ for the strata; the subsequent resonance test provides the sharper identification. The best-recovered channel, $\Psi$, is also the most consequential: the
perturbed field and its topology derive from $\Psi$ alone, so every island chain, field-line trace, and resonance check below rests on it.

\paragraph{Performance by stratum.}
Table~\ref{tab:strata} breaks the headline metrics down by mode number, inferred $m$, Lundquist number $S$, and normalized pressure $\beta_0$. The forecaster columns use the 358 curated validation rows. Field error is the median physical-gauge relative $L_2$ error, and $\gamma$ RMSE is in $\log_{10}\gamma$ (dex), both for the four-member ensemble. The classifier columns use the 1{,}344 unstable inputs of the full validation population, and recall is the unstable-case recall at the conservative operating point. The two populations differ, so the counts within a row are not comparable. The $S$ and $\beta_0$ bins are tertiles of the curated validation rows ($\log_{10}S$ edges 6.72 and 7.59; $\beta_0$ edges $2.8\times10^{-5}$ and $9.5\times10^{-5}$). The same edges are applied to the classifier inputs, so their counts per bin are unequal. Classifier inputs carry an inferred $m$ only where truth fields exist (346 of the 1{,}344 unstable inputs).

In every stratum with more than six rows, field error stays between 0.26 and 0.33, and $\gamma$ RMSE between 0.07 and 0.11~dex. The six-row $n{=}3$ stratum has a median absolute $\gamma$ error of 0.032~dex, so its 0.244~dex RMSE comes from a small number of rows. Of the three Lundquist-number tertiles, the lowest is the hardest for both outputs. The curated validation and test sets contain no $n{=}4$ rows and only six and eight $n{=}3$ rows, so field and $\gamma$ accuracy is established for $n{=}1$ and $2$ (about 98\% of curated rows); for $n{=}3$ and $4$ only the screening result is supported.

The single validation miss is an $n{=}3$ input in the high-$S$ and high-$\beta_0$ tertiles. The operating point was selected on validation, so validation recall is not an independent estimate. On the held-out test population (run once), recall is 100\% across all strata (0 of 1{,}258 missed).

 \begin{table}[h]
    \caption{Performance by stratum on validation. Forecaster columns use the
358 curated rows and classifier columns use the 1{,}344 unstable inputs. Metric definitions and bin edges are given in the text.}
    \label{tab:strata}
    \centering
    \small
    \begin{tabular}{lrccrc}
      \toprule
      Stratum & Rows & Field error & $\gamma$ RMSE & Unstable & Recall\\
      \midrule
      All & 358 & 0.285 & 0.087 & 1{,}344 & 99.93\%\\
      \midrule
      $n{=}1$ & 267 & 0.265 & 0.085 & 636 & 100\%\\
      $n{=}2$ & 85 & 0.332 & 0.071 & 397 & 100\%\\
      $n{=}3$ & 6 & 0.247 & 0.244 & 187 & 99.47\%\\
      $n{=}4$ & 0 & --- & --- & 124 & 100\%\\
      \midrule
      Inferred $m{=}1$ & 222 & 0.263 & 0.079 & 216 & 100\%\\
      Inferred $m{=}2$ & 136 & 0.288 & 0.100 & 130 & 100\%\\
      \midrule
      $\log_{10}S$ low & 119 & 0.326 & 0.106 & 609 & 100\%\\
      $\log_{10}S$ mid & 118 & 0.277 & 0.074 & 436 & 100\%\\
      $\log_{10}S$ high & 121 & 0.272 & 0.079 & 299 & 99.67\%\\
      \midrule
      $\beta_0$ low & 119 & 0.265 & 0.094 & 174 & 100\%\\
      $\beta_0$ mid & 119 & 0.308 & 0.068 & 337 & 100\%\\
      $\beta_0$ high & 120 & 0.287 & 0.097 & 833 & 99.88\%\\
      \bottomrule
    \end{tabular}
  \end{table}

\begin{figure}[t]
  \centering
  \includegraphics[width=0.85\linewidth]{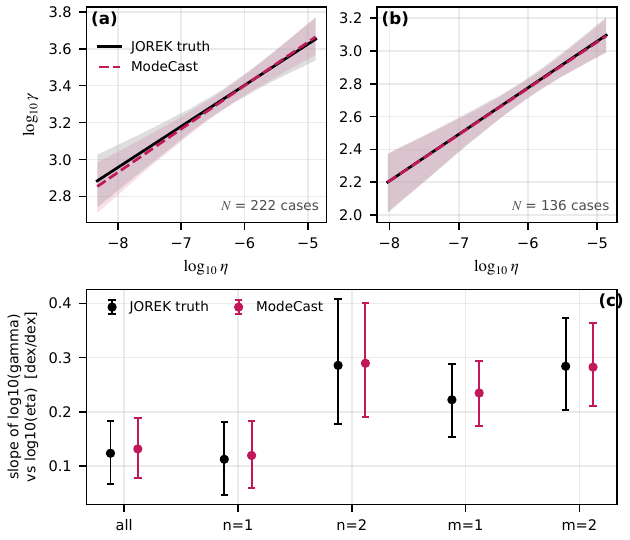}
  \caption{Resistivity-trend fidelity: fitted marginal trends of
  $\log_{10}\gamma$ against $\log_{10}\eta$ on identical rows, truth against prediction with 95\% bands (inferred $m{=}1$: 222 cases; $m{=}2$: 136), and the fitted slope per stratum with 95\% intervals (the six-row $n{=}3$ stratum is omitted). Slopes are marginal associations across an LHS design; no analytic exponent is claimed.}
  \label{fig:etatrend}
\end{figure}

\paragraph{Controlled slices and parameter-trend fidelity.}
These checks use the physics that the sampling design permits. First, within-equilibrium $\gamma(n)$ slices, the one controlled variable an LHS design contains: each equilibrium is simulated at several mode numbers with everything else fixed, and geometry-disjoint splits leave complete slices in validation. Over the 55 validation equilibria with $\ge 2$ unstable harmonics (53 with two, 2 with three; cluster bootstrap over equilibria; permutation chance $\approx 0.49$ on every rate): top-harmonic agreement 1.000, exact ordering 0.982, and median $|\Delta\hat\gamma - \Delta\gamma|$ 0.056 dex against the standard-practice twin's 0.101 (the twin also orders well, 0.964 top; the recipe's advantage is increment-magnitude fidelity; Table~\ref{tab:nslice}). Second, resistivity-trend fidelity on identical rows, by stratum (inferred $m{=}1$: 222 rows; $m{=}2$: 136). The recipe$-$truth $\eta$-coefficient difference is CI-consistent with zero in every stratum (marginal $+0.008$ over all rows; the 21-parameter partial agrees; Figure~\ref{fig:etatrend}), while the twin underestimates the coefficient on the inferred $m{=}1$ stratum, CI-separated (marginal $-0.032$; partial agrees) and matches elsewhere. No analytic exponent is claimed: marginal trends in an LHS design are not partial derivatives. Third, the measured layer-width trend: the FWHM of the $|J_\phi|$ chord at the resonant surface, extracted identically from truth and prediction on all 358 validation rows, gives $\log_{10}\delta$ against $\log_{10}S$ slopes of $-0.059$ (truth) and $-0.069$ (prediction), paired difference CI-consistent with zero (Figure~\ref{fig:scaling}b) --- a field-level check no scalar surrogate can attempt. The measured FWHM proxy is far shallower than the asymptotic layer model because the width floors near the grid cell at high $S$, consistent with the resolution reading above. Fourth, counterfactual sweeps: with the equilibrium and every other input held fixed, resistivity alone is swept $\pm 1$ decade through the trained model (15 base cases per stratum; $\eta$-derived scalars recomputed; builder validated at every base point; the gate routes no synthetic point stable). The responses are smooth and cluster by stratum (medians $+0.49$ inferred $m{=}2$, $+0.53$ inferred $m{=}1$), but the intervals overlap, and no ground truth exists at synthetic points, so this serves as a robustness probe only. Certifying the analytic $\eta^{1/3}$ (internal kink) and $\eta^{3/5}$ (tearing) exponents \citep{coppi1976kink, hastie1987kinkscaling, furth1963tearing} takes controlled fixed-geometry resistivity scans, which this dataset does not contain. Section~\ref{sec:transfer} scores the surrogate on an external scan that does, and reports how far the reference simulations there sit from the analytic exponents.

\paragraph{Resonance localization and parity.}
Two further structural checks read the layer physics off the predictions directly (truth and ensemble; the twin's stored dump carries only $\gamma$, so it has no field columns here). Localization: resistive modes localize at their resonant surface, and the radial position of the predicted $|J_\phi|$ peak selects the correct $q{=}m/n$ crossing in 357 of 358 validation cases, among a median of 4 candidate surfaces per inferred $m{=}1$ row and 7 per inferred $m{=}2$ row. Its median offset from the truth's resonant surface is 0.67 cells, compared with the truth's 0.66, with a median predicted-to-truth peak offset of 0.00 cells. The rational-surface proximity map is an input, so tight localization alone could echo the features; the load-bearing number is the correct-surface selection, which the map does not supply: it marks every candidate without naming the resonant one. Parity: the score $P$ measures the symmetry of the perturbed flux across the rational surface; a tearing-type mode is even there (constant-$\psi$, flux peaking at the surface) while an internal kink is odd (flux passing through zero). On a gauge-aligned $\pm 8$-cell chord, the truth's strata separate cleanly (inferred $m{=}2$ median $+0.68$, the constant-$\psi$ even signature, against $+0.27$ for inferred $m{=}1$, CI-separated; a $\pm 5$-cell window gives the same verdict). The ensemble carries the signature: $m{=}2$ median $+0.66$ with parity-sign agreement 1.000 and median parity error 0.027, $m{=}1$ $+0.25$ at 0.928. In shaped geometry, the inferred-$m{=}1$ parity is not odd-dominated, so the discriminator the data supports is the strata's parity gap, and the prediction reproduces it.

\begin{figure}[t]
  \centering
  \includegraphics[width=0.82\linewidth]{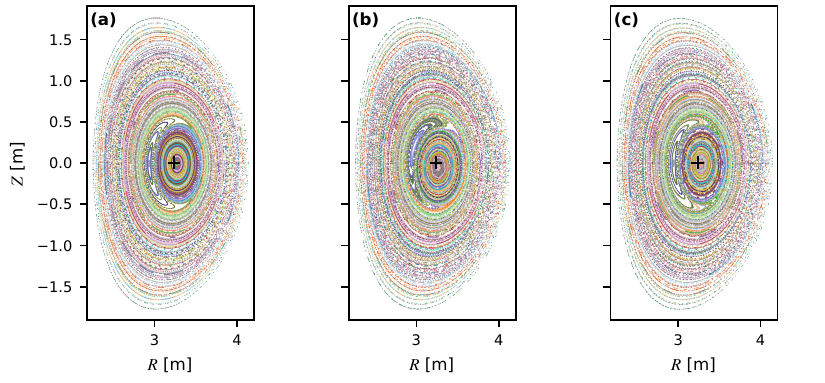}
  \caption{Poincar\'e sections turn predicted eigenmodes into magnetic
  topology: island position and width are set by the relative phases across fields, the gauge results in a physical form. (a) JOREK truth, (b) a conventionally trained twin (ViT: bare inputs, per-channel loss), (c) \modelname, each rotated onto truth by its joint gauge phase. The twin does not reproduce the truth's $(1,1)$ island chain (puncture-density total variation 11.5\% vs 6.7\%); the phase-blind ablation, which displaces rather than destroys the chain, is in Appendix~\ref{app:poincare} (validation internal kink).}
  \label{fig:poincare}
\end{figure}

\begin{table}[t]
  \caption{Growth rate ordering within equilibria: 55 held-out equilibria
  with at least two unstable harmonics, cluster bootstrap over equilibria; chance from within-slice permutation. Spacing error is the median $|\Delta\log_{10}\hat\gamma - \Delta\log_{10}\gamma|$ between harmonics. The single ordering miss is one $k{=}3$ slice whose predicted $n{=}3$ drop is absent; all 80 growing slices reproduce the pattern.}
  \label{tab:nslice}
  \centering
  \small
  \begin{tabular}{lccc}
    \toprule
    & Top harmonic & Full ordering & Spacing error (dex)\\
    \midrule
    \modelname\ (ensemble) & \textbf{1.000} & \textbf{0.982} & \textbf{0.056}\\
    Bare-input U-Net & 0.964 & 0.945 & 0.101\\
    Chance (permutation) & 0.49 & 0.49 & ---\\
    \bottomrule
  \end{tabular}
\end{table}

\FloatBarrier
\section{System comparisons, ablations, and cost}
\label{app:systems}

\begin{table}[h]
  \caption{Cross-pool generalization between the small (curated) and large
  datasets: the same recipe trained on each, evaluated on both validation splits (validation only, no test scores; single models; physical-gauge relative $L_2$ median; own normalization per model; geometry-disjointedness verified for every pairing). Each column is one evaluation split, so cells compare down columns; the diagonal is each model on its home split, and the off-diagonal measures transfer. Transfer is asymmetric: leaving home costs the small-dataset model $+0.115$ (0.304 to 0.419) and the large-dataset model $+0.020$ (0.288 to 0.309), a roughly six-fold smaller shift.}
  \label{tab:crosspool}
  \centering
  \small
  \begin{tabular}{lcc}
    \toprule
    & Eval: small (curated, 358) & Eval: large (growing, 1{,}344)\\
    \midrule
    Trained on small (2{,}718) & 0.304 & 0.419\\ %
    Trained on large (9{,}968) & 0.309 & 0.288\\ %
    \bottomrule
  \end{tabular}
\end{table}

\paragraph{Screening and false-alarm accounting.}
A forecaster trained only on growing modes fabricates growth from stable inputs, so screening is a separate model trained on the complete labeled population (36{,}734 simulations, including stable cases). The gate is the maximum of a gradient-boosted tree model over 43 scalar features and a small three-seed convolutional ensemble (ROC AUC $\approx 0.999$ on validation).
At the conservative operating point (Table~\ref{tab:routing}), zero unstable cases are missed --- 0 of 1{,}258 on the once-run test and 0 of 4{,}569 composed inputs at frozen thresholds --- at a per-configuration false-alarm rate of 19.1\% (an equilibrium flagged if any of its four mode numbers is; 9.45\% per input). Zero misses is a finite-sample result: with 0 of 1{,}258, the one-sided 95\% upper bound on the miss rate is 0.24\% (exact Clopper--Pearson, $1-0.05^{1/1258}$; the rule of three, $3/1{,}258$, agrees to two digits), and validation has one miss in 1{,}344 (recall 99.93\%, Table~\ref{tab:routing}). Never certifying an unstable configuration as stable costs roughly one unnecessary check in five configurations. In the resulting hybrid loop, stable inputs are cleared after the classifier with a calibrated probability and a growth rate bound, flagged inputs receive the full eigenmode forecast, and undecided inputs abstain and reroute to JOREK.

\begin{table}[h]
  \caption{Three-way outcome of the stability classifier on its validation
  population (4{,}731 inputs, 1{,}344 unstable) at the conservative operating point, with the once-run test line.}
  \label{tab:routing}
  \centering
  \small
  \begin{tabular}{lccl}
    \toprule
    Outcome & Share & Content & \\
    \midrule
    Resolved stable (surrogate) & 64.8\% & 1 unstable case missed of 1{,}344 & \\
    Abstained & 6.9\% & 8.0\% truly unstable & \\
    Forecast (surrogate) & 28.3\% & 98.4\% truly unstable & \\
    \midrule
    \multicolumn{4}{l}{Unstable-case recall 99.93\% (1 of 1{,}344); accounting above.}\\
    \multicolumn{4}{l}{Held-out test (run once): 0 of 1{,}258 unstable missed; 95\% upper bound 0.24\%.}\\
    \bottomrule
  \end{tabular}
\end{table}

The per-configuration rate is the use-level number, since screening a configuration queries all its mode numbers, and a false alarm costs one unnecessary check while a miss invalidates a stability conclusion. A second, zero-abstention operating point on the same coverage--risk curve forecasts every routed input directly rather than abstaining; both are disclosed, and the conservative one is used throughout the paper.

\begin{table}[h]
  \caption{Within-system ablations (validation; format as
  Table~\ref{tab:main}; ---$^\dagger$ validation only, no test score). The conventional rows of
  Table~\ref{tab:main} and the UPT, CALM-PDE, and two-stage rows here
  (marked $\dagger$) are scored on validation only: they postdate the single test run, which was not repeated. Backbone swap: external architectures receive the full system, parameter-matched into the incumbent's band (three-seed medians; CALM-PDE one raw-weight seed). Component rows: one component removed. External seed spread (ViT 0.364--0.399) is comparable to the backbone gaps; removals are several times larger. Backbone block ranked bold/underline; component rows unranked. The single-model row's four ensemble members are four seeds of one recipe; spread 0.304 / 0.317 / 0.334 (min / median / max).}
  \label{tab:ablation}
  \centering
  \resizebox{\linewidth}{!}{%
  \begin{tabular}{lcccccc}
    \toprule
    Variant & Params & Rel.\ $L_2$ (physical) & (per-channel) & $\gamma$ $R^2$ & $\gamma$ RMSE & Test\\
    \midrule
    \multicolumn{7}{l}{\emph{Backbone swap under the full system}}\\
    U-Net (= \modelname\ single) & 145M & \textbf{0.304} & \textbf{0.252} & \textbf{0.968} & \textbf{0.091} & \textbf{0.322}\\
    FNO & 143M & 0.385 & 0.330 & 0.960 & 0.102 & \underline{0.396}\\ %
    ViT & 156M & 0.399 & 0.336 & 0.947 & 0.118 & 0.407\\ %
    UPT & 190M & \underline{0.353} & \underline{0.302} & \underline{0.963} & \underline{0.099} & ---$^\dagger$\\ %
    CALM-PDE & 177M & 0.391 & 0.338 & 0.939 & 0.127 & ---$^\dagger$\\ %
    \midrule
    \multicolumn{7}{l}{\emph{Component sensitivity (same network, one component removed)}}\\
    No physics features & 145M & 0.665 & 0.554 & 0.878 & 0.178 & 0.677\\
    Phase-blind loss & 145M & 0.863 & 0.250 & 0.966 & 0.094 & 0.861\\
    Two-stage variant & 145M & 0.315 & 0.267 & 0.967 & 0.093 & ---$^\dagger$\\
    \bottomrule
  \end{tabular}}
\end{table}

\begin{table}[h]
  \caption{Per-channel ensemble error under the two metrics (SSIM higher is
  better; test rows from the stored once-run predictions). Smooth channels sit at 5--14\% relative $L_2$; the resistive-layer channels hold the residual in the thin sheet at the rational surface (Appendix~\ref{app:results}). Anchor rows: the \emph{true} field displaced by one grid cell, showing each metric's response to sheet placement.}
  \label{tab:channels}
  \centering
  \small
  \begin{tabular}{lccccccc}
    \toprule
    & $\Psi$ & $u$ & $J_\phi$ & $\omega$ & $\rho$ & $T$ & $v_\parallel$\\
    \midrule
    Relative $L_2$ (validation) & 0.052 & 0.078 & 0.129 & 0.268 & 0.323 & 0.182 & 0.376\\
    Relative $L_2$ (test, run once) & 0.055 & 0.078 & 0.141 & 0.265 & 0.353 & 0.184 & 0.396\\
    SSIM (validation) & 0.966 & 0.978 & 0.984 & 0.980 & 0.905 & 0.908 & 0.935\\ %
    SSIM (test, stored) & 0.965 & 0.978 & 0.983 & 0.979 & 0.891 & 0.905 & 0.924\\ %
    \midrule
    Anchor: one-cell-shifted truth, rel.\ $L_2$ & 0.058 & 0.236 & 0.589 & 0.991 & 0.371 & 0.123 & 0.595\\ %
    Anchor: one-cell-shifted truth, SSIM & 0.989 & 0.969 & 0.941 & 0.918 & 0.944 & 0.974 & 0.945\\ %
    \bottomrule
  \end{tabular}
\end{table}

\begin{table}[h]
  \caption{Training-data equivalent of the physics features: the factor of
  additional data the poorer set needs to match the richer one over the shared error range; growth rate factors are seed-informed pooled-slope fits (the $\gamma$ ladders are parallel across sets).}
  \label{tab:prices}
  \centering
  \small
  \begin{tabular}{lcc}
    \toprule
    Feature set added & Mode structure & Growth rate\\
    \midrule
    $q$-profile maps & 2.6--5.8$\times$ & 1.5$\times$\\ %
    + rational surfaces & 3.7--4.0$\times$ & 1.1--1.3$\times$\\ %
    + similarity parameters & 0.8--1.7$\times$ (no gain) & 2.0$\times$\\ %
    \midrule
    all, over bare & no shared error range & 3.6--3.8$\times$\\ %
    \bottomrule
  \end{tabular}
\end{table}

\paragraph{Unfiltered-selection counterfactual.}
Before the large dataset was fixed as the full-scale set, an \emph{unfiltered} selection added 7{,}003 stable shape-leaf rows to the same 9{,}968 large-dataset training pairs (21{,}300 runs in all, mixed 2{,}200-case split). Phase-level probes had already measured stable-run fields as seed-floor noise with no forecastable content, so the unfiltered selection is kept only as a counterfactual for training on the complete population. Table~\ref{tab:counterfactual} reports it: the backbone ordering matches the small dataset's, and the mixed-trained model transfers to the small dataset at parity (0.291 against the specialist's 0.304 on identical rows). Re-scored on the shared growing split, every mixed-trained cell trails the growing-only model at any size (best mixed 0.328 against 0.288), so stable-row admixture rather than volume capped these arms.

\begin{table}[h]
  \caption{Counterfactual: the full recipe and full-system backbone swaps trained on the unfiltered selection (16{,}971 pairs including stable shape rows), evaluated on its own mixed 2{,}200-case split --- levels comparable only within this table. ViT ran below the parameter-matching band (32.9M). Best per column bold, second underlined; ---$^\dagger$ validation only, no test score.}
  \label{tab:counterfactual}
  \centering
  \resizebox{\linewidth}{!}{%
  \begin{tabular}{lcccccc}
    \toprule
    Method & Params & Rel.\ $L_2$ (physical) & (per-channel) & $\gamma$ $R^2$ & $\gamma$ RMSE & Test\\
    \midrule
    FNO (full system) & --- & 0.544 & 0.463 & \underline{0.978} & \underline{0.074} & ---$^\dagger$\\ %
    ViT (full system) & 32.9M & \underline{0.512} & \underline{0.421} & 0.965 & 0.094 & ---$^\dagger$\\ %
    \textbf{\modelname\ (single, from scratch)} & 145M & \textbf{0.463} & \textbf{0.381} & \textbf{0.984} & \textbf{0.063} & ---$^\dagger$\\ %
    \bottomrule
  \end{tabular}}
\end{table}

\begin{table}[h]
  \caption{Capabilities and per-input cost of the main approaches to linear MHD stability analysis. Rollout field surrogates forecast nonlinear field evolution (reduced-MHD blob simulations, MAST camera data) rather than performing stability analysis; they are included as the nearest field-prediction lineage. Cost is wall-clock per input (one equilibrium, one toroidal mode number). The \modelname\ figure is the median latency at batch size 1 of the full pipeline from the design parameters on one NVIDIA RTX PRO 6000 (fp32): the equilibrium network, the physics features and the single large-dataset forecaster (Appendix~\ref{app:e2e}); 2.3~s on one CPU core. JOREK figures are campaign medians per linear run (1.3--4.3~h on 32--64 cores, 83--137 core-hours), to which the equilibrium solve adds about 2~min.}
  \label{tab:capabilities}
  \centering
  \resizebox{\linewidth}{!}{%
  \begin{tabular}{lccccr}
    \toprule
    Method & Stability & Growth rate & Mode structure & Uncertainty & Cost\\
    \midrule
    Linear MHD simulation~\citep{hoelzl2021jorek} & \cmark & \cmark & \cmark & --- & 1.3--4.3~h (32--64 cores)\\
    Tabular regressors~\citep{piccione2020pb, xu2024dtm} & \cmark & \cmark & \xmark & \xmark & $\sim$\textmu s\\
    Rollout field surrogates~\citep{gopakumar2023fno} & \xmark & \xmark & fields over time & \xmark & $\sim$ms (GPU)\\
    \textbf{\modelname\ (this work)} & calibrated & with intervals & one-shot eigenmode & abstention &  0.05~s (one GPU) / 2.3~s (1 core)\\ %
    \bottomrule
  \end{tabular}}
\end{table}

\FloatBarrier
\section{Additional visualizations}
\label{app:gallery}

Cases in this appendix are selected by recorded rules rather than by eye: gallery rows are the nearest-rank 25th, 50th, and 75th percentiles of the joint physical-gauge error within each mode number. The small dataset's validation split contains no unstable $n{=}4$ case, so the gallery spans $n = 1$--$3$.

\begin{figure}[p]
  \centering
  \includegraphics[width=\linewidth]{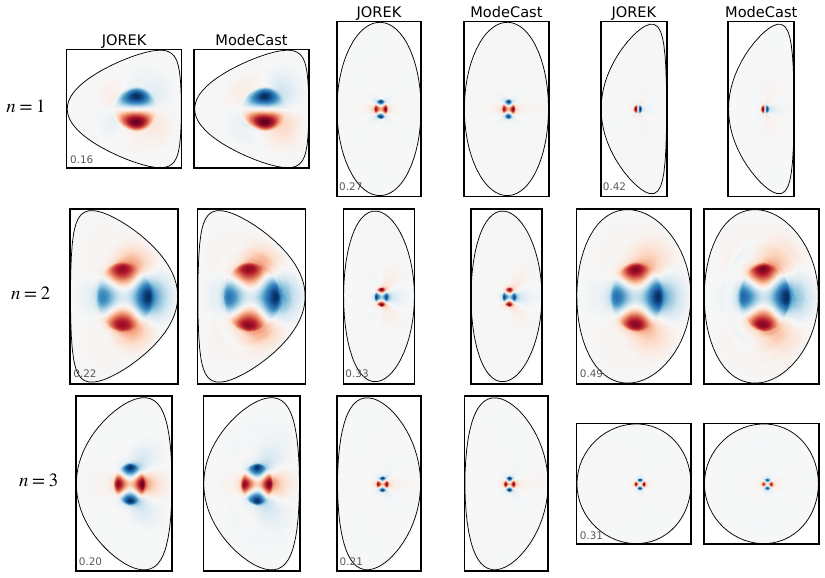}
  \caption{Mode gallery: $\mathrm{Re}(\hat\Psi)$, truth beside prediction,
  for the 25th / 50th / 75th error percentiles (left to right) within each toroidal mode number $n = 1$--$3$ (top to bottom). Annotations give the joint physical-gauge relative $L_2$.}
  \label{fig:gallery}
\end{figure}

\begin{figure}[p]
  \centering
  \includegraphics[width=\linewidth]{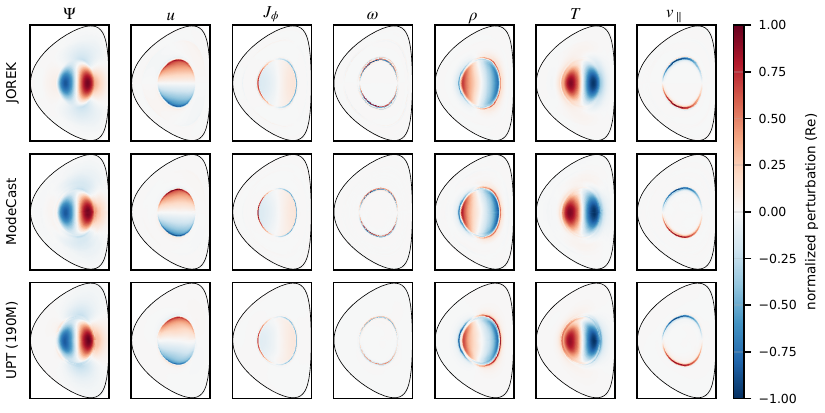}
  \caption{Seven-channel comparison against the strongest external
  backbone under the full system in the case of Figure~\ref{fig:showcase}: the parameter-matched UPT (190M) at its best landed seed, rendered from the stored checkpoint; table rows report the median seed.}
  \label{fig:baseline7}
\end{figure}

Figure~\ref{fig:gallery} shows $\mathrm{Re}(\hat\Psi)$, truth beside prediction, across mode numbers and difficulty; the 90th-percentile joint error (0.637) is more than twice the median, so these tails, rather than a uniform blur, dominate the aggregate. Figure~\ref{fig:baseline7} compares all seven channels against the strongest external backbone under the full system (UPT) in the case of Figure~\ref{fig:showcase}; the externals reproduce the smooth channels and lose the resistive-layer structure ($\omega$, $v_\parallel$) first.

\section{Poincar\'e methodology}
\label{app:poincare}

\paragraph{Field-line tracing.}
The perturbed poloidal flux at toroidal angle $\varphi$ is $\psi(R, Z,\varphi) = \psi_{\mathrm{eq}} + \alpha\,\mathrm{Re}[\hat{\Psi} e^{in\varphi}]$, with the perturbation scaled to $\alpha = 1.5\%$ of the equilibrium flux span --- inside the linear regime, chosen so island structure is visible at print size, and identical for every model so the comparison is amplitude-matched. With $B_R = -({1}/{R})\,\partial_Z \psi$, $B_Z = ({1}/{R})\,\partial_R \psi$ and $B_\varphi = F_0/R$, field lines integrate as $dR/d\varphi = -({R}/{F_0})\,\partial_Z\psi$ and $dZ/d\varphi = ({R}/{F_0})\,\partial_R\psi$. We launch 230 seeds (inboard and outboard midplane plus a vertical fan through the axis, which samples the $m{=}1$ island in any orientation) and record 280 toroidal transits per seed, plotting punctures at $\varphi = 0$ with one color per field line.

\begin{figure}[tp]
  \centering
  \includegraphics[width=\linewidth]{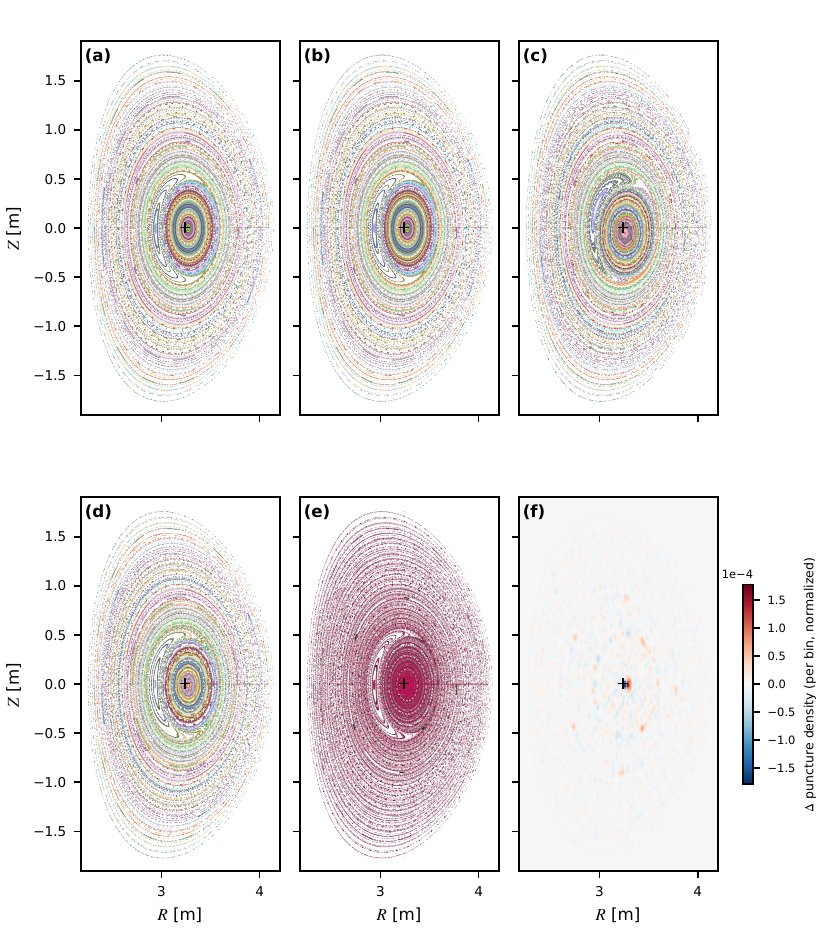}
  \caption{Full Poincar\'e comparison for the case of
  Figure~\ref{fig:poincare}: (a) truth as stored; (b) truth in the training gauge (coincident here); (c) control and (d) ensemble, gauge-rotated; (e) truth--ensemble overlay; (f) puncture-density difference, total variation 6.7\% smoothed (35.5\% raw) vs 10.1\% (38.5\%) for the control.}
  \label{fig:poincare6}
\end{figure}

\paragraph{Gauge alignment.}
An isolated linear mode is defined up to one global phase, which at fixed $\varphi$ acts as a rigid rotation of the mode pattern, so predictions must be brought to a common gauge before topology is compared. Each model is rotated onto the truth by its single best-fit phase over all seven channels jointly (the physical mode gauge of Section~\ref{sec:method}). Aligning on $\Psi$ alone would instead absorb each model's $\Psi$ misrotation and thereby remove exactly the inter-channel phase error under test, since the trace uses only $\Psi$. After the joint rotation, the ensemble's $\Psi$ lies $0.2^\circ$ from the truth while the phase-blind control retains a ${\sim}20^\circ$ residual, displacing its island chain (Figure~\ref{fig:poincare6}). For this case, the stored truth already coincides with the training gauge (canonical phase $0.2^\circ$), so panels (a) and (b) of Figure~\ref{fig:poincare6} coincide.

\paragraph{Puncture-density total variation.}
Punctures are binned on a $220^2$ grid over the section and normalized to a probability density; the reported distance is the total variation $\tfrac{1}{2}\sum_{\mathrm{bins}}|p - q|$. In raw histograms, the finite-sampling speckle of 230$\times$280 punctures dominates: truth vs.\ ensemble 35.5\%, truth vs.\ control 38.5\%, truth vs.\ the conventionally trained twin 40.7\%. Smoothing both densities with a Gaussian of 1.4 bins before the distance suppresses the speckle and exposes the structural disagreement: 6.7\% for the ensemble against 11.5\% for the twin (Figure~\ref{fig:poincare}) and 10.1\% for the phase-blind control (Figure~\ref{fig:poincare6}).

\FloatBarrier
\section{Dataset details}
\label{app:dataset}
The dataset derives from six simulation campaigns, four on cloud infrastructure (32 cores per run, median wall-clock 3.5--4.3~h) and two on an HPC cluster (64 cores per run, median 1.3~h), totaling about 6.2 million CPU core-hours over 59{,}886 completed runs, or 83--137 core-hours per linear run:

\begin{table}[h]
  \caption{Simulation campaigns behind the dataset.}
  \label{tab:campaigns}
  \centering
  \small
  \begin{tabular}{lccr}
    \toprule
    Campaign & Completed runs & Cores/run & Core-hours\\
    \midrule
    Set 1: Uniform sampled scan, growth-biased & 810 & 32 & 96{,}825\\
    Set 2: LHS scan, growth-biased & 10{,}251 & 32 & 1{,}163{,}862\\
    Set 3: LHS scan, growth-biased & 8{,}888 & 32 & 1{,}180{,}461\\
    Set 4: LHS scan, unbiased & 9{,}765 & 32 & 1{,}384{,}539\\
    Set 5: LHS scan, growth-biased & 19{,}722 & 64 & 1{,}528{,}393\\
    Set 6: Dense sampling around resonant modes  & 10{,}450 & 64 & 855{,}516\\
    \bottomrule
  \end{tabular}
\end{table}

  \begin{table}[h]
    \caption{Composition of each split by toroidal mode number $n$ and
    inferred poloidal number $m$ (Appendix~\ref{app:physics}) for the curated forecaster dataset, and by $n$ for the classifier population (unstable / all inputs). Splits are geometry-disjoint.}
    \label{tab:composition}
    \centering
    \small
    \begin{tabular}{llrrr}
      \toprule
      & Stratum & Train & Validation & Test\\
      \midrule
      \multicolumn{5}{l}{\emph{Forecaster (curated dataset)}}\\
      $n{=}1$ & inferred $m{=}1$ & 1{,}687 & 221 & 221\\
              & inferred $m{=}2$ & 310 & 46 & 39\\
      $n{=}2$ & inferred $m{=}1$ & 4 & 1 & 0\\
              & inferred $m{=}2$ & 671 & 84 & 87\\
      $n{=}3$ & inferred $m{=}2$ & 37 & 6 & 8\\
      $n{=}1,2,4$ & inferred $m{=}3$--$6$ & 9 & 0 & 0\\
      Total & & 2{,}718 & 358 & 355\\
      \midrule
      \multicolumn{5}{l}{\emph{Classifier (unstable / all inputs)}}\\
      $n{=}1$ & & 4{,}840 / 7{,}666 & 636 / 1{,}000 & 607 / 948\\
      $n{=}2$ & & 2{,}985 / 9{,}134 & 397 / 1{,}172 & 369 / 1{,}137\\
      $n{=}3$ & & 1{,}273 / 9{,}919 & 187 / 1{,}282 & 167 / 1{,}240\\
      $n{=}4$ & & 870 / 10{,}015 & 124 / 1{,}277 & 115 / 1{,}244\\
      Total & & 9{,}968 / 36{,}734 & 1{,}344 / 4{,}731 & 1{,}258 /
  4{,}569\\
      \bottomrule
    \end{tabular}
  \end{table}

\paragraph{Labeling and splits.}
Each completed run is labeled from its linear-phase time trace: an exponential fit to the perturbation energy gives $\log_{10}\gamma$ and the fit's $r^2$. Since the perturbation energy scales as $E \propto e^{2\gamma t}$, the growth rate is the halved log-energy slope, $\gamma = \tfrac{1}{2}\, \mathrm{d}\ln E/\mathrm{d}t$, so that $\gamma$ is the amplitude rate throughout. Runs whose perturbation never grows are labeled stable (retained: 33{,}464 stable, 10{,}235 unstable, and 2{,}335 marginally unstable pairs, sampled across 21 design parameters). The fit procedure was audited against synthetic traces with known rates: at five or more e-foldings inside the fit window, the label error is $\sigma \leq 0.016$ in $\log_{10}\gamma$ with negligible bias, and surviving weak-class runs carry a median of 16 e-foldings, so growth rate label noise (0.006 median) sits far below the model's residual. The halving convention is checked separately by refitting $\gamma$ to ground-truth energies and comparing it against the growth rate that JOREK reports for the same run. All splits are geometry-disjoint (an equilibrium's geometry hash appears in exactly one of train/validation/test: 36{,}734 / 4{,}731 / 4{,}569 pairs, verified by hash intersection at build time); the classifier trains and evaluates on this complete population, stable cases included. Two training sets are drawn from the pool: the \emph{small dataset} keeps runs with clean exponential fits ($r^2 \ge 0.99$; 2{,}718 train, 358 validation, 355 test) and the $\sim$3.7$\times$ larger \emph{large dataset} relaxes the fit floor (9{,}968 train; 1{,}344 held out); levels compare within a dataset, not across, and every test column is run once. The small dataset's validation split contains no unstable $n{=}4$ case, so per-mode claims cover $n \le 3$. The large dataset's relaxed fit floor changes what a label means: a snapshot stored before one mode dominates is a superposition whose weights depend on the perturbation seed rather than the equilibrium, so part of that shape target is unpredictable in principle, and the $\gamma$ target is masked where the fit supports no value (1{,}332 of the 1{,}344 held-out cases carry one).

\paragraph{Normalization.}
Fields are resampled from the flux-aligned simulation mesh to a common $128{\times}128$ poloidal grid; one campaign was archived natively at $151{\times}101$ and, uniquely, with every fifth checkpoint saved. Fields are z-scored per channel, with statistics computed only on each training selection; each trained model carries its own normalization, and predictions are always compared in physical units. Growth rates are handled as $\log_{10}\gamma$ throughout, matching their four-decade dynamic range.

\section{Full pipeline: from design parameters to stability}
\label{app:e2e}

A JOREK stability assessment has two stages: an equilibrium solve turns the design parameters into an equilibrium (about 2~min on 32 cores), and the linear run turns that equilibrium into the fastest-growing mode and its growth rate (1.3--4.3~h). \modelname\ replaces the second stage. In our campaigns, every equilibrium is fixed by 21 design parameters drawn by Latin hypercube: the plasma shape, the toroidal field, and the coefficients of the current, density, and temperature profiles, together with resistivity and viscosity, which enter only the linear run. The first stage is therefore a deterministic map that a network can learn as well. Placed in front of \modelname, it gives the full pipeline from the design parameters to the eigenmode and growth rate.

\paragraph{Method.}
The first network is a U-Net with the same architecture as \modelname's. It takes the 21 design parameters, broadcast as constant channels on the flux-aligned $128\times128$ grid, and predicts the equilibrium maps that \modelname\ uses: $\Psi$, $J_\phi$, $\rho$, $T$, $R$, $Z$, and $q$. It is trained on the equilibria of the training split; no validation or test equilibrium is seen. The physics features are computed from the predicted equilibrium as from a solved one, and the second network is the single large-dataset \modelname\ of Table~\ref{tab:main}, unchanged.

\paragraph{Accuracy.}
The predicted equilibrium matches the solved one to 0.2--1.7\% (median relative $L_2$ per map on validation), and the set of rational surfaces inside the plasma is exactly right in 98\% of validation and 99\% of test cases. Table~\ref{tab:e2e} compares the full pipeline with the same \modelname\ fed with JOREK's equilibrium. From the design parameters alone, the growth rate is as accurate as from the solved equilibrium, and the mode error rises by about 0.04. The remaining gap comes from small shifts of the rational surfaces: the resistive layer of the mode sits on them and is only a few cells wide, so a sub-cell shift of a surface moves the sharpest part of the mode (Section~\ref{sec:results}).

\begin{table}[h]
  \caption{Full pipeline from the design parameters (curated benchmark; relative $L_2$ medians; $\gamma$ RMSE in $\log_{10}$). Both rows use the single large-dataset \modelname\ of Table~\ref{tab:main}; they differ only in the equilibrium it receives.}
  \label{tab:e2e}
  \centering
  \small
  \begin{tabular}{llcccc}
    \toprule
    Split & Equilibrium & Rel.\ $L_2$ (physical) & (per-channel) & $\gamma$ $R^2$ & $\gamma$ RMSE\\
    \midrule
    Validation & JOREK & 0.309 & 0.261 & 0.976 & 0.079\\
               & Predicted (full pipeline) & 0.343 & 0.298 & 0.976 & 0.079\\
    \midrule
    Test & JOREK & 0.318 & 0.272 & 0.981 & 0.071\\
         & Predicted (full pipeline) & 0.356 & 0.297 & 0.980 & 0.072\\
    \bottomrule
  \end{tabular}
\end{table}

\paragraph{Cost.}
On one NVIDIA RTX PRO 6000 GPU (fp32, batch 1), the full pipeline takes 53~ms per design and mode number (2.3~s on one CPU thread): about 27~ms for each network and under 1~ms for the physics features. The equilibrium is computed once per design, so all four mode numbers of a design take 60~ms. JOREK needs the equilibrium solve and then one linear run per mode number, so the full pipeline is about $10^5\times$ faster for a single mode number and up to $10^6\times$ for all four mode numbers of a design. The first network learns the equilibria of these campaigns and applies within their parameter ranges; equilibria from other sources go to \modelname\ directly.

\end{document}